\documentclass[journal]{IEEEtran}

\usepackage{amssymb}
\usepackage{bbm}
\usepackage{mathbbol}
\usepackage{newtxmath}
\usepackage{amsmath}
\usepackage{graphicx}
\usepackage{subfigure}
\usepackage{caption}
\usepackage{booktabs}
\usepackage{algorithmic}
\usepackage[top=0.65in, bottom=0.65in, left=0.7in, right=0.7in]{geometry}
\usepackage{bm}
\usepackage{array}    
\usepackage{wrapfig}   

\usepackage{multirow}

\usepackage[linesnumbered,ruled,vlined]{algorithm2e}

\usepackage{subcaption}          
\usepackage{xcolor} 

\usepackage{cite}     
\usepackage{hyperref}

\begin{document}

\title{%Daul-Branch Vector Quantized-Aided Satellite Digital Semantic Communication for High-Resolution RSI With AFDM Transmission
%Dual-Branch Vector-Quantization-Aided Satellite Digital Semantic Communication for High-Resolution RSI Over AFDM Channels
%Dual-Branch Vector-Quantization-Aided Satellite Digital Semantic Communication for High-Resolution RSI Over AFDM Channels
Dual-Branch Vector-Quantization-Aided Satellite Digital Semantic Communication with Index Compression for High-Resolution RSI Over AFDM
}

\author{Jianqiao Chen, Nan Ma, \textit{Member, IEEE}, Xiaodong Xu, \textit{Senior Member, IEEE},
    Tingting Zhu, \textit{Member, IEEE}, Huishi Song, Chen Dong, Rui Meng, \textit{Member, IEEE}, Wenkai Liu, Ke Peng and Ping Zhang, \textit{Fellow, IEEE}
    % <-this % stops a space
\thanks{
%This work was supported in part by the National Key R$\&$D Program of China under Grant 2025YFF0514703; in part by the 6G Key Technologies R$\&$D and Ecosystem Cultivation project. 
%The work of Rui Meng was supported by the China Postdoctoral Science Foundation. 
%(\textit{Corresponding author: Nan Ma.})

%Jianqiao Chen, Nan Ma, Xiaodong Xu, Tingting Zhu and Huishi Song are with the ZGC Institute of Ubiquitous-X Innovation and Applications, and Beijing Key Laboratory of 6G DOICT converged and Cloud-Native Mobile Information Networks, Beijing 100876, China. 
%Nan Ma, Xiaodong Xu and Tingting Zhu are also with the State Key Laboratory of Networking and Switching Technology, Beijing University of Posts and Telecommunications, Beijing 100876, China (e-mail: \{chenjianqiao; zhutingting; songhuishi\}@zgc-xnet.com. \{manan; xuxiaodong\}@bupt.edu.cn).

%Chen Dong, Rui Meng, Wenkai Liu, Ke Peng, and Ping Zhang are with the State Key Laboratory of Networking and Switching Technology, Beijing University of Posts and Telecommunications, Beijing 100876, China (e-mail: \{dongchen; buptmengrui; liuwenkai; pengke; pzhang\}@bupt.edu.cn). 

Jianqiao Chen, Tingting Zhu and Huishi Song are with the ZGC Institute of Ubiquitous-X Innovation and Applications, and Beijing Key Laboratory of 6G DOICT converged and Cloud-Native Mobile Information Networks, Beijing 100876, China. 
Nan Ma, Xiaodong Xu, Chen Dong, Rui Meng, Wenkai Liu, Ke Peng, and Ping Zhang are with the State Key Laboratory of Networking and Switching Technology, Beijing University of Posts and Telecommunications, Beijing 100876, China (e-mail: \{chenjianqiao; zhutingting; songhuishi\}@zgc-xnet.com. \{manan; xuxiaodong; dongchen; buptmengrui; liuwenkai; pengke; pzhang\}@bupt.edu.cn). (\textit{Corresponding author: Nan Ma.})

}% <-this % stops a space
}

\maketitle
\thispagestyle{empty}
\begin{abstract}

High-resolution remote sensing imagery (RSI) transmission is constrained by satellite-ground bandwidth and channel impairments, yet existing methods struggle to simultaneously achieve extreme compression and robust transmission. To address this, we propose a dual-branch vector-quantization-aided satellite digital semantic communication (DVQ-SDSC) framework for RSI transmission over affine frequency division multiplexing (AFDM), whose bandwidth savings arise from two interrelated aspects. First, at the source-coding level, a dual-branch framework is developed to unify deep joint semantic coding, VQ-aided index transmission, channel estimation and adaption in an end-to-end architecture; 
departing from symmetric encoder designs, the codec is recast as an asymmetric dual-branch architecture that separately processes the high-frequency residuals and the low-frequency structural semantics, with gated fusion and channel-adaptive reconstruction jointly restoring the semantic content.
Second, at the index-coding level, we develop a principal component analysis (PCA)-aided codebook reordering to align index topology with latent correlations, and devise group differential pulse-code modulation (G-DPCM) to encode prediction residuals rather than absolute indices, lowering the index bitrate while locally isolating clipping and channel errors. A two-stage training strategy further decouples channel impairments from the semantic codec. FAIR1M experiments over 3GPP NTN-TDL-D demonstrate that DVQ-SDSC with G-DPCM index coding outperforms the conventional JPEG-LDPC scheme at the base rate of 0.0625 bits per pixel (BPP), and that G-DPCM applies directly to the trained codec without retraining, yielding additional index compression at no extra cost.

\end{abstract}

\begin{IEEEkeywords}
Satellite communication, digital semantic communication, vector quantization, AFDM, index compression.
\end{IEEEkeywords}

\IEEEpeerreviewmaketitle

\section{Introduction}

\subsection{Background and Significance}

\IEEEPARstart{s}{atellite} communication has gained increasing attention owing to its large-scale connectivity and seamless wireless coverage, while high-quality remote sensing imagery (RSI)
transmission is vital for environmental monitoring, including the urban planning, agricultural management, and disaster response [1]. However, the explosive growth of low Earth orbit (LEO)-satellite deployments and their massive high-resolution imagery, combined with severely limited satellite bandwidth, make efficient and timely transmission of such big data increasingly urgent [2]. Additionally, high-velocity LEO-satellite motion causes rapid channel variation, degrading signal quality and elevating the bit error rate, thereby compromising communication reliability [3]. 
Conventional Shannon-theoretic schemes compress all bits uniformly and employ separate source-channel coding (SSCC), thus performing poorly in bandwidth-constrained, highly interfered satellite-to-ground image links [4], [5].

Rapidly evolving semantic communication has emerged as a breakthrough paradigm for enhancing transmission efficiency [6], [7]. By leveraging deep joint source-channel coding (DJSCC), it synergistically unifies source and channel coding within a single optimization framework, which enhances noise resilience and effectively mitigates the \textit{cliff effect} [8], [9]. Various DJSCC schemes have been developed for different communication source modes [10], [11], [12] and physical layer transmission modes [13], [14], [15]. 
However, despite these advances, existing semantic communication schemes still face three major limitations in high-resolution RSI transmission over dynamic LEO-satellite channels.
\textit{Limitation 1:} Most existing DJSCC schemes rely on analog or end-to-end continuous representations, lacking a practical discrete semantic mapping mechanism compatible with standardized digital communication infrastructures.
\textit{Limitation 2:} 
Conventional single-branch semantic codecs process low-frequency (LF) structural content and high-frequency (HF) texture within a shared representation, prioritizing the former over the latter under extreme bandwidth constraints.
\textit{Limitation 3:} Prevailing schemes transmit semantic information and optimize under idealized, perfect channel state information (CSI), which is highly mismatched with the rapidly time-varying LEO-satellite channels.
We next summarize the current state of research and present the contributions of this work.

\subsection{Related Works and Contributions}

As to the processing of RSI, recent research has explored the potential of semantic communication systems in evolving satellite communications [16], [17]. Jiang et al. [18] propose a framework that combines feature extraction and semantic source-channel coding, achieving excellent accuracy in the classification task. Cao et al. [19] applies segmentation model enhancement for fine-grained semantic segmentation, which prioritizes semantically vital content via task-driven selection and real-time CSI and uses an signal-to-noise ratio (SNR)-aware stack-based channel codec for adaptive coding. As to RSI compression and transmission tasks, a generative foundation model-based semantic satellite communication framework is proposed to safeguard key features and enhance visual quality [20]. Leveraging known channels and prior images, it refines performance while reducing frequent bandwidth adjustments. Tan et al. [21] propose to compress RSIs into semantic information and residuals for high-quality reconstruction under limited bandwidth, adapt semantic preferences via an attention feature module across SNRs, and enable variable semantic lengths through a variance-based position mask module. As to internet of everything scenarios, a semantic communication framework is developed to adapt to heterogeneous receiver capacities, which uses Swin Transformer model to transmit task-relevant features for reducing bandwidth [22]. 
To tackle bandwidth limitations and semantic knowledge base (SKB) vulnerability, Lin et al. [23] propose a spatially attention-assisted joint coding-modulation framework for semantic communication, improving semantic precision and SKB-based data recovery while ensuring consistency. 
%via adversarial training.

Vector-quantized (VQ)-enabled semantic communication represents semantic features via trainable codebooks, whose corresponding discrete indices can be directly mappable to standard digital modulation. It thus integrates semantic communication into existing satellite communication infrastructure naturally with low hardware cost and full compatibility [24], [25]. An adaptive semantic communication framework is proposed by integrating a deep reinforcement learning agent with a vector-quantized variational autoencoder (VQ-VAE) [26]. It dynamically selects modulation via real-time Doppler and delay spread, while receiver-side post-equalization reduces residual frequency offsets. To improve transmission efficiency and link stability in free space optical (FSO)-based RSIs transmission, Chen et al. in [27] propose an FSO-semantic communication scheme using VQ-VAE with spatial normalization for preserving image details. A semantic forwarding-based framework is proposed for satellite-terrestrial networks that jointly optimizes the semantic encoder and codebook to shape constellation symbols [28]. The feature-wise linear modulation-based channel-aware reconstruction and codebook split-enhanced model division multiple access enhance robustness and spectral efficiency.

Although VQ-based RSI processing has undergone preliminary research, the conventional VQ architecture still constrains its applicability to high-resolution RSI, particularly in semantic compression and reconstruction. Moreover, the above models have not been trained or evaluated under LEO-satellite channel characteristics.
Chen et al. in [29] combines semantic communication with orthogonal time-frequency space (OTFS) modulation to mitigate Doppler and boost efficiency and integrates adversarial training with 3GPP channel model. However, it adopts traditional channel encoder and decoder for enhancing error correction capability. 
%Compared with OTFS, the recently developed affine frequency division multiplexing (AFDM) offers a lower-complexity one-dimensional discrete affine Fourier transform (DAFT) structure instead of a two-dimensional delay-Doppler transform, requires only a one-dimensional pilot guard for channel estimation, and provides comparable bit-error performance with reduced pilot overhead []. 
In contrast to OTFS, the recently proposed affine frequency division multiplexing (AFDM) employs a lower-complexity one-dimensional discrete affine Fourier transform (DAFT) structure rather than a two-dimensional delay-Doppler transform, which therefore requires only one-dimensional pilot guards for channel estimation and attains comparable bit-error performance with reduced pilot overhead [30], [31].
Several AFDM-based communication schemes for LEO-satellite systems have developed recently [32], [33]. However, research combined with semantic communication is still quite lacking. 

\begin{figure*}[t]
\centering
\resizebox{0.86\textwidth}{0.18\textheight}{\includegraphics{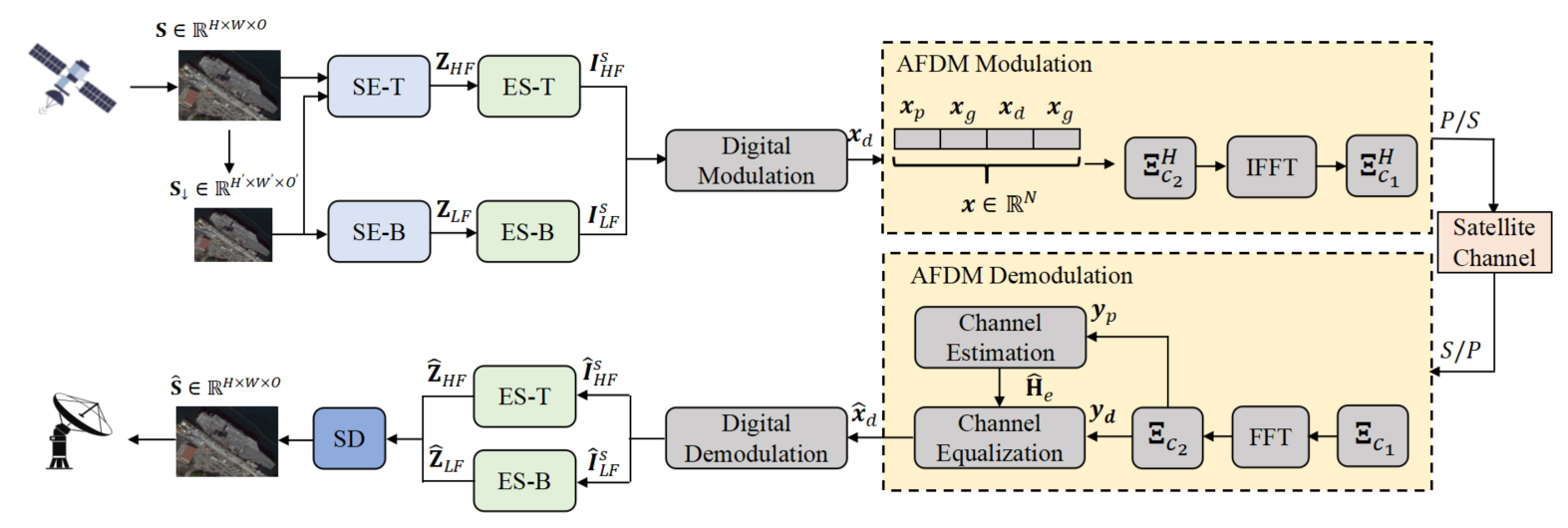}}
\captionsetup{font={footnotesize}}
\caption{Overall architecture of DVQ-SDSC framework over AFDM-based satellite channels.}
\label{fig1}
\end{figure*}

To overcome these limitations, we propose the dual-branch vector-quantization-aided satellite digital semantic communication (DVQ-SDSC) framework for RSI transmission over AFDM-based satellite channels. It unifies deep joint semantic coding, VQ-based discrete mapping, residual-indexed transmission, sparse channel estimation, and SNR-adaptive decoding within a single end-to-end architecture, systematically addressing (i) the lack of a discrete semantic mapping (\textit{Limitation 1}), (ii) the separation of LF structural and HF residual semantics at source-coding level (\textit{Limitation 2}), and (iii) SNR-adaptive reconstruction over estimated CSI (\textit{Limitation 3}). The key contributions are summarized as follows:

\begin{itemize}
\item[]
\hspace{-0.5cm} $\bullet$
\hspace{0.1cm}\textit{Dual-Branch Framework with LF/HF Extraction and Codebook Mapping:} 
We propose a dual-branch framework for semantic feature extraction and VQ quantization mapping, tailored to high-resolution RSI transmission. Unlike single-branch codecs that entangle structure and texture, the encoder separately extracts HF texture details and LF structural semantics, while the decoder performs gated fusion and reconstruction of the two branches, preserving both under limited bandwidth. To the best of our knowledge, this is the first work to combine asymmetric LF/HF semantic branching with VQ-based discrete mapping for high-resolution RSI transmission over satellite channels.

\hspace{-0.5cm} $\bullet$
\hspace{0.1cm}\textit{Asymmetric LF/HF Semantic Branching:} 
We depart from single-branch encoders that represent structure and texture within a shared latent and instead devise an asymmetric two-branch architecture in which each branch is quantized through a dedicated VQ codebook. The top branch extracts HF residuals via bicubic subtraction and refines them through multi-scale fusion and channel attention before quantization; the bottom branch encodes LF structural semantics at a reduced spatial resolution. Their outputs are reconciled by a gated fusion module through channel-wise adaptive weighting and residual refinement, yielding the final reconstruction.

\hspace{-0.5cm} $\bullet$
\hspace{0.1cm}\textit{Correlation-based Index Compression and Transmission:} 
Conventional VQ-based architectures transmit indices using fixed-length codes, leaving adjacent-index correlation unexploited. To address this, we propose a principal component analysis (PCA)-aided codebook reordering that aligns the index topology with latent feature correlations, and develop group-wise differential pulse-code modulation (G-DPCM), which encodes prediction residuals rather than absolute indices and thereby reduces the average bitrate. Group-wise error isolation confines clipping and channel-induced errors locally, preventing distortion propagation under channel fading.

\hspace{-0.5cm} $\bullet$
\hspace{0.1cm}\textit{Experimental Validation and Performance Evaluation:} 
We develop a two-stage training strategy that isolates channel-induced impairments from the semantic codec. Extensive experiments on FAIR1M over 3GPP NTN-TDL-D channels demonstrate that DVQ-SDSC with G-DPCM index coding outperforms conventional Photographic Experts Group (JPEG) + low-density parity-check (LDPC) in peak signal-to-noise ratio (PSNR), multiscale structural similarity (MS-SSIM), and learned perceptual image patch similarity (LPIPS) at 0.0625 bits per pixel (BPP). Furthermore, G-DPCM reduces the index bitrate relative to fixed-length coding without codec retraining or additional transmission overhead.

\end{itemize}

The remainder of this paper is organized as follows. Section II presents the system model and DVQ-SDSC framework for high-resolution RSI transmission over AFDM-based satellite channels. Section III details the dual-branch asymmetric codec, VQ mapping, and joint optimization strategy. 
Section IV elaborates on the proposed PCA-aided codebook reordering and G-DPCM for index compression and transmission. Section V reports experimental results and performance evaluations. Finally, Section VI concludes the paper.

\textit{Notation:} Boldface small letters denote vectors and boldface capital letters denote matrices. 
$\mathbb{R}$, $\mathbb{Z}$ and $\mathbb{C}$ denote the real number field, integer number field, and complex number field, respectively. 
$[\cdot]^{T}$ denotes matrix transpose. $[\cdot]^{H}$ denotes conjugate transpose.
$\|\cdot\|_{2}$ denotes Euclidean norm.

\section{DVQ-SDSC Framework for high-resolution RSI Transmission over AFDM-based channels}

As shown in Fig. 1, the DVQ-SDSC framework for high-resolution RSI transmission over AFDM-based satellite channels adopts a dual-branch asymmetric encoder to decouple HF texture details from LF structural semantics. The underlying rationale is twofold: LF semantics alone are insufficient for photorealistic reconstruction under extreme compression, while HF features transmitted without LF priors lead to structural drift and visual artifacts.

\subsection{Semantic Joint Source-Channel Encoder at the Satellite}

Let the original high-resolution image dataset be $\mathcal{D}$, with each image  $\mathbf{S} \in \mathbb{R}^{H\times W \times O} \subset \mathcal{D}$, where $H$, $W$ and $O$ are the height, width, and channel size of each image, respectively. The low-resolution image $\mathbf{S}_{\downarrow} \in \mathbb{R}^{H^{\prime}\times W^{\prime} \times O^{\prime}}$ related to $\mathbf{S}$ can be obtained via bicubic interpolation. 
At the transmitter, two dedicated semantic encoders (SEs) are developed to extract features from the high-resolution and low-resolution images, referred to as the top semantic encoder (SE-T) and bottom semantic encoder (SE-B), respectively. 
The extracted HF and LF semantic features are denoted by $\mathbf{\bm{Z}}_{HF} = \left[\bm{z}_{0},..., \bm{z}_{k_{z}},..., \bm{z}_{K_{z}-1} \right]^{T} \in \mathbb{R}^{K_{z} \times M_{z}}$ and $\mathbf{\bm{Z}}_{LF} = [\bm{z}^{\prime}_{0},..., \bm{z}^{\prime}_{k^{\prime}_{z}},..., \bm{z}^{\prime}_{K^{\prime}_{z}-1} ]^{T} \in \mathbb{R}^{K^{\prime}_{z} \times M^{\prime}_{z}}$, respectively, which can be calculated as 
\begin{equation}
\mathbf{\bm{Z}}_{HF} = f_{SE-T}\left(\mathbf{\bm{S}}, \mathbf{S}_{\downarrow}, \bm{\alpha}_{T} \right), \mathbf{\bm{Z}}_{LF} = f_{SE-B}\left(\mathbf{S}_{\downarrow}, \bm{\alpha}_{B} \right),
\setcounter{equation}{1}
\label{eq1}
\end{equation}
where $f_{SE-T}\left(\cdot, \bm{\alpha}_{T}\right)$ and $f_{SE-B}\left(\cdot,  \bm{\alpha}_{B}\right)$ denote the SE-T and SE-B with parameters $\bm{\alpha}_{T}$ and $\bm{\alpha}_{B}$, respectively. 

The dual-branch quantized codebooks are assisted by embedding spaces for the top semantic feature (ES-T) and bottom semantic feature (ES-B), which correspond respectively to the HF features and the LF features. Let the ES-T and ES-B be $\mathbf{\bm{C}}_{HF} = \left[\bm{c}_{0},..., \bm{c}_{k_{T}},..., \bm{c}_{K_{T}-1} \right]^{T} \in \mathbb{R}^{K_{T} \times N_{T}}$ and $\mathbf{\bm{C}}_{LF} = [\bm{c}^{\prime}_{0},..., \bm{c}^{\prime}_{k_{B}},..., \bm{c}^{\prime}_{K_{B}-1} ]^{T} \in \mathbb{R}^{K_{B} \times N_{B}}$, where $K_{T}\left(K_{B}\right)$ denotes the codebook size, and each element $\bm{c}_{k_{T}} \in \mathbb{R}^{N_{T}}(\bm{c}^{\prime}_{k_{B}} \in \mathbb{R}^{N_{B}} )$ is called a semantic codeword with dimension size $N_{T}\left(N_{B} \right)$ and index $k_{T}\left( k_{B}\right)$. 
%The ES-T and ES-B serve as the shared semantic knowledge bases for both the transmitter and receiver. 
%Integrated with the semantic encoder and semantic decoder, it facilitates the joint optimization of semantic feature extraction, quantization, and image reconstruction. 
The feature elements in $\mathbf{\bm{Z}}_{HF}$ and $\mathbf{\bm{Z}}_{LF}$ are passed through discretization bottleneck and generates the mapping index with respect to the ES-T and ES-B as follows:
\begin{align}
\hat{\bm{z}}_{k_{z}}=q\left(\bm{z}_{k_{z}} \right) \triangleq \operatorname*{argmin}_{\bm{c}_{j}} \left\| \bm{z}_{k_{z}} - \bm{c}_{j} \right\|_{2}, j=1,2,...,K_{T}, \nonumber\\
\hat{\bm{z}}^{\prime}_{k_{z}^{\prime}}=q(\bm{z}^{\prime}_{k_{z}^{\prime}}) \triangleq \operatorname*{argmin}_{\bm{c}^{\prime}_{j}} \| \bm{z}^{\prime}_{k^{\prime}_{z}} - \bm{c}^{\prime}_{j} \|_{2}, j=1,2,...,K_{B},
\setcounter{equation}{1}
\label{eq2}
\end{align}
where $q\left(\cdot \right)$ denotes quantization operator, and $\bm{c}_{j}( \bm{c}^{\prime}_{j})$ denotes the element with the $j$th index of ES-T (ES-B). 

After all the semantic features are mapped into the discrete space, the corresponding index sequences of semantic features of $\mathbf{\bm{Z}}_{HF}$ 
and $\mathbf{\bm{Z}}_{LF}$ are obtained, 
which are denoted as $\bm{I}^{s}=\left\{ \bm{I}^{s}_{HF}, \bm{I}^{s}_{LF} \right\} \in \mathbb{Z}^{\left({K_{z} + K^{\prime}_{z}}\right)}$. 
Through \textit{Digital Modulation} module, $\bm{I}^{s}$ is mapped into the transmitted symbol vector $\bm{x}_{d} \in \mathbb{C}^{L_{s}} $, where $L_{s}$ denotes the length of modulated symbols.

\subsection{AFDM-Based Transmission Under LEO-Satellite Channel}

To combat the high-mobility LEO-satellite communications with time-varying channels and large Doppler, we consider the AFDM-based multicarrier waveform. 
The processing procedure of AFDM-based transmission is shown on the right side of Fig. 1. Let $N$ denote the number of AFDM chirp subcarries. Each AFDM frame spans a duration of $T=N\cdot \triangle t$. The sampling rate is given by $\frac{1}{\triangle t}$, and the subcarrier spacing to $\triangle f=\frac{1}{T}$. 
Let $\bm{x}_{p} \in \mathbb{C}^{N_{p}}$ denote pilot sequence and $\bm{x}_{g} \in \mathbb{C}^{N_{g}}$ denote guard sequence. The data sequence, pilot sequence, and guard sequence are combined to form the DAFT symbol vector $\bm{x}=\left[x_{1},...,x_{m},...,x_{N}    \right]^{T} \in \mathbb{C}^{N}$. The $x_{m}$ is transformed into time-domain symbol $s_{n}$ through an $N$-point IDAFT as
\begin{equation}
s_{n} = \sum_{m=0}^{N-1}x_{m}\phi_{n}\left(m\right), n=0,1,...,N-1,
\setcounter{equation}{3}
\label{eq3}
\end{equation}
where $\phi_{n}\left(m\right) = \frac{1}{\sqrt{N}} e ^ {j2\pi \left(c_{1}n^{2} + c_{2}m^{2} + \frac{nm}{N} \right)}$, 
$c_{1}$ and $c_{2}$ denote chirp parameters for AFDM time-frequency resource division.

To make the channel lie in a periodic domain, a chirp periodic prefix (CPP) should be added to the modulated signal, i.e., $s_{n} = s_{N+n}e^{-j2\pi c_{1}\left(N^{2} + 2Nn \right)}, n=-L_{cp},...,-1$, where $L_{cp}$ denotes any integer greater than or equal to the value in samples of the maximum delay spread of the wireless channel. 
From \eqref{eq3}, its vector form can be given by  
\begin{equation}
\bm{s} = \mathbf{A}^{H}\bm{x}=\mathbf{\Xi}_{c_{1}}^{H}\mathbf{F}^{H}\mathbf{\Xi}_{c_{2}}^{H}\bm{x},
\setcounter{equation}{4}
\label{eq4}
\end{equation}
where $\mathbf{A} \triangleq \mathbf{\Xi}_{c_{2}}\mathbf{F}\mathbf{\Xi}_{c_{1}}$ denotes the DAFT matrix, $\mathbf{\Xi}_{c} \triangleq \operatorname{diag} (e^{-j2\pi cn^{2}}, n=0,1,...,N-1 )$, and $\mathbf{F}$ denotes the discrete Fourier transform (DFT) matrix with entries $\frac{1}{\sqrt{N}}e^{-j2\pi mn/N}$.

The ground-space LEO-satellite channel can be characterized in delay-Doppler (DD) domain as 
\begin{equation}
g_{n}\left(l \right) = \sum_{i=1}^{L_{p}} h_{i}e^{-j2\pi f_{i}n}\delta\left(l-l_{i} \right),
\setcounter{equation}{5}
\label{eq5}
\end{equation}
where $L_{p}$ denotes the number of multi-paths, $\delta\left(\cdot \right)$ denotes the Dirac delta operator, and $h_{i}$, $f_{i}$ and $l_{i}$ denote the complex gain, Doppler shift, and integer delay associated with the $i$th path, respectively. Let $\nu_{i} \triangleq Nf_{i}=  \bar{\nu}_{i}+\iota_{i}$, where $\nu_{i} \in \left[-\nu_{max}, \nu_{max}\right]$ denotes the normalized Doppler shift with respect to the subcarrier spacing, $\bar{\nu}_{i}$ denotes its integer part whereas $\iota_{i}$ denotes the fractional part satisfying $-\frac{1}{2}\leq \iota_{i} \leq \frac{1}{2}$. 

After the parallel to serial conversion, transmission over the channel and CPP discarding, the received symbols in DAFT-domain can be calculated as
\begin{equation}
\bm{y} = \sum_{i=1}^{L_{p}}h_{i}\mathbf{H}_{i}\bm{x} + \mathbf{A}\bm{w} = \mathbf{H}_{e}\bm{x} + \widetilde{\bm{w}}, 
\setcounter{equation}{6}
\label{eq6}
\end{equation}
where $\bm{w}=\left[0,...,w_{n},...,w_{N-1}\right]^{T}$ with $w_{n} \sim \mathcal{CN}\left(0, 1\right)$ denoting the additive Gaussian noise, $\mathbf{H}_{i}=\mathbf{A}\mathbf{\Gamma}_{i} \mathbf{\Delta}_{\epsilon_{i}}\mathbf{\Pi}_{l_{i}}\mathbf{A}^{H} $, $\mathbf{\Delta}_{\epsilon_{i}}=\operatorname{diag}(e^{-j2\pi \frac{\epsilon_{i}}{N}n }, n=0,1,...,N-1)$,  $\mathbf{\Pi}_{l_{i}}$ denotes the forward cyclic-shift matrix in terms of $l_{i}$, and $\mathbf{\Gamma}_{i}$ denotes a $N \times N$ diagonal matrix, i.e., $\mathbf{\Gamma}_{i} = \operatorname{diag}\left(    \begin{cases} e^{-j2\pi c_{1}\left(N^{2} - 2N\left(l_{i} - n \right) \right)}, & n \leq l_{i} \\1, & n \ge l_{i}\end{cases}, n=0,...,N-1\right)$.
%which can be expressed as 
%\begin{equation}
%\mathbf{\Gamma}_{i} = \operatorname{diag}\left(    \begin{cases} e^{-j2\pi c_{1}\left(N^{2} - 2N\left(l_{i} - n \right) \right)}, & n \leq l_{i} \\1, & n \ge l_{i}\end{cases}, n=0,...,N-1\right).
%\setcounter{equation}{8}
%\end{equation}

\subsection{Semantic Joint Source-Channel Decoder at UEs}

After quantization and modulation processing, the index sequences related to semantic features are transmitted over the AFDM link. The \textit{Channel Estimation} module first computes an estimate of effective channel matrix $\widehat{\mathbf{H}}_{e} \in \mathbb{C}^{N \times N}$ based on pilot and received symbols. The \textit{Channel Equalization} module then performs to recover transmitted signals, i.e., $\widehat{\bm{x}}_{d} = \widehat{\mathbf{H}}_{e}^{-1}\bm{y}_{d}$, where $\bm{y}_{d}$ denotes the received symbols for channel estimation. The equalized data symbol $\widehat{\bm{x}}_{d}$ is finally passed to \textit{Digital Demodulation} module, which reverses the modulation mapping to obtain the recovered index vector $\widehat{\bm{I}}^{s}=\{ \widehat{\bm{I}}^{s}_{HF}, \widehat{\bm{I}}^{s}_{LF} \}$. Based on $\widehat{\bm{I}}^{s}$ and feature embedding spaces (ES-T/ES-B), the HF features $\widehat{\mathbf{\bm{Z}}}_{HF}$ and LF features $\widehat{\mathbf{\bm{Z}}}_{LF}$ can be mapped, 
which are passed to the \textit{Semantic Decoder} (SD) to reconstruct the original image as 
\begin{equation}
\widehat{\mathbf{S}} = f_{SD}(\{\widehat{\mathbf{\bm{Z}}}_{HF}, \widehat{\mathbf{\bm{Z}}}_{LF}\}, \bm{\beta} ),
\setcounter{equation}{7}
\label{eq7}
\end{equation}
where $f_{SD}\left(\cdot, \bm{\beta} \right)$ denotes the SD module with parameter $\bm{\beta}$.

\subsection{Problem Formulation}

For transmitting high-resolution RSIs in satellite-to-ground channel with very limited bandwidth resources, we formulate the DVQ-SDSC framework as a joint optimization of dual-branch-aided semantic extraction, VQ mapping and reconstruction with AFDM-based transmission as follows:
\begin{align}
&\hspace{-0.3cm}\left(\bm{\alpha}^{*}_{T}, \bm{\alpha}^{*}_{B}, \bm{\beta}^{*}, \mathbf{\bm{C}}^{*}_{HF},\mathbf{\bm{C}}^{*}_{LF} \right)  \nonumber\\
&= \operatorname*{argmin}_{\bm{\alpha}_{T}, \bm{\alpha}_{B}, \bm{\beta}, \mathbf{\bm{C}}_{HF},\mathbf{\bm{C}}_{LF}} \mathbb{E}_{p(\sigma)}\mathbb{E}_{p(\bm{h})}\mathbb{E}_{p(\mathbf{S}, \hat{\mathbf{S}})}[d(\mathbf{S}, \hat{\mathbf{S}} )  ],
\setcounter{equation}{7}
\label{eq8}
\end{align}
where $\mathbb{E}_{p(x)}[\cdot]$ denotes the expectation with respect to the distribution $p(x)$, $p(\sigma)$ and $p(\bm{h})$ model the statistical distribution of fading channel and noise, respectively, $p(\mathbf{S}, \hat{\mathbf{S}})$ denotes the joint distribution of original and reconstructed semantic features, and $d(\cdot)$ denotes a perceptual distortion metric. 
The optimization variable $\bm{\alpha}^{*}_{T}, \bm{\alpha}^{*}_{B}, \bm{\beta}^{*}, \mathbf{\bm{C}}^{*}_{HF},\mathbf{\bm{C}}^{*}_{LF}$ collectively govern the dual-branch configurations, VQ mapping and channel adaption strategies, jointly minimizing the perceptual distortion over AFDM-based satellite channels. 
Furthermore, VQ-index transmission provides an additional avenue for bandwidth reduction, owing to its compact discrete-index representation of semantic features. On the other hand, the achievable rate reduction remains suboptimal due to the use of conventional fixed-length index coding.

\section{Design of DVQ-SDSC Module}

This section details the major DVQ-SDSC modules, which employ lightweight neural network architectures tailored to the asymmetric satellite-to-ground computational capabilities.

\begin{figure}
\centering
\subfigure[]{
  \label{fig:subfig:a}
   \includegraphics[width = 0.43\textwidth]{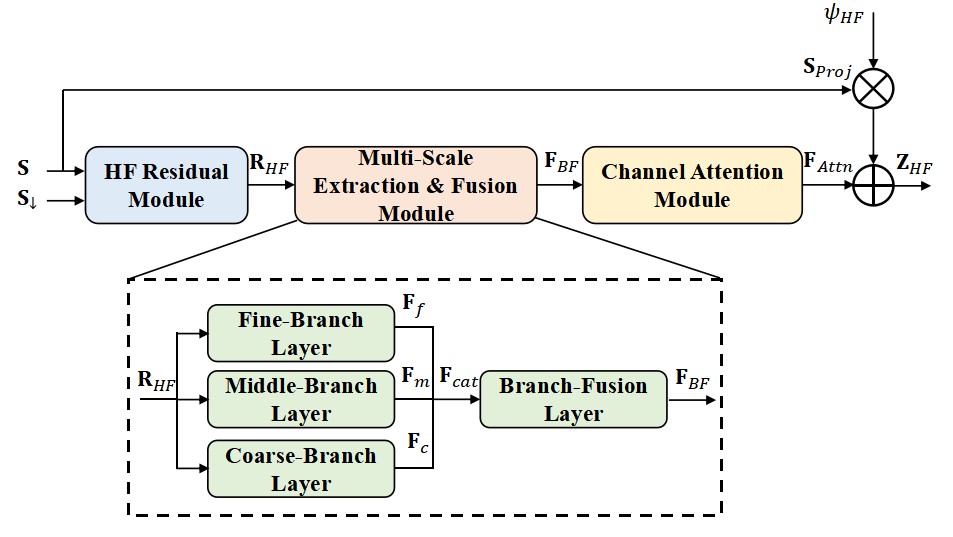}}
\hspace{1in}
\subfigure []{
 \label{fig:subfig:b}
 \includegraphics[width = 0.43\textwidth]{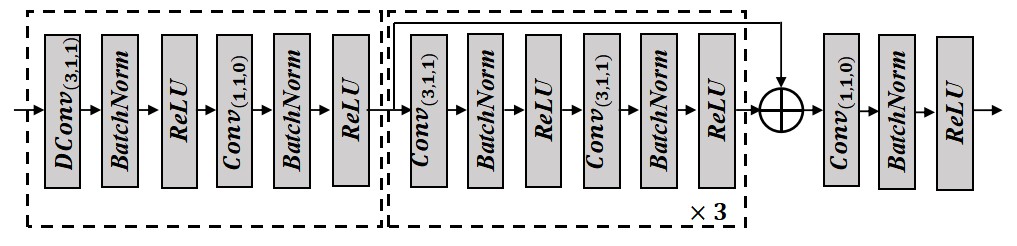}}
\captionsetup{font={footnotesize}}
\caption{
Architecture of the proposed SE-T HF feature extraction network. (a) Overall pipeline. (b) Detailed internal structure of each parallel branch. $DConv_{(k,s,p)}$ denotes the depth-wise separable convolution operation with kernel-size $k$, stride-size $s$ and padding-size $p$. $Conv_{(k,s,p)}$ denotes the convolution operation. $BatchNorm$ denotes batch normalization operation. $ReLU$ denotes rectified linear unit function. 
} 
\label{fig:subfig}
\end{figure}

\subsection{SE-T Design for HF Feature Extraction}

As shown in Fig. 2(a), the SE-T is designed to capture discriminative HF residuals (i.e., edges, texture, and fine-grained details) from HR images, compensating for detail loss in the LR encoded features while suppressing noise, thereby providing informative priors for cross-frequency fusion.

$\bullet$ \textbf{\textit{HF Residual} Module}:  
The SE-T first explicitly computes the HF residual by subtracting the bicubic-up-sampled low-resolution image from the original high-resolution input, thereby directly isolating the target HF components to be reconstructed. 
The HF residual information captured by \textit{HF Residual} module is calculated as  
\begin{equation}
\mathbf{R}_{HF} = \mathbf{\bm{S}} - U_{b}\left(\mathbf{S}_{\downarrow} \right),
\setcounter{equation}{9}
\label{eq9}
\end{equation}
where $U_{b}\left(\cdot \right)$ denotes the bicubic up-sampling operator that restores $\mathbf{S}_{\downarrow}$ to the spatial resolution of $\mathbf{\bm{S}}$. By leveraging the bicubic interpolation as the baseline, the residual term focuses the network's capacity on reconstructing non-linear HF details rather than relearning LF content.

$\bullet$ \textbf{\textit{Multi-scale Extraction}$\And$\textit{Fusion} Module}: To capture HF patterns spanning fine-grained edges, medium-scale textures, and coarse structural layouts, the \textit{Multi-scale Extraction}$\And$\textit{Fusion} module employs three parallel branches, namely \textit{Fine-Branch}, \textit{Middle-Branch}, and \textit{Coarse-Branch} layers that share an identical architecture, followed by a \textit{Branch-Fusion} layer. As illustrated in Fig. 2(b), each branch consists of a down-sampling stage (left, dashed box) and a customized residual block for high-order feature extraction (right, dashed box).

As to \textit{Fine-Branch} layer, it targets thin edges and micro-textures by applying a minimal down-sampling stride, preserving spatial resolution to retain fine-grained details. 
As to \textit{Middle-Branch} layer, it focuses on medium-scale textures by adopting two consecutive down-sampling operations which reduces spatial redundancy while maintaining semantic context. 
As to \textit{Coarse-Branch} layer, it captures large-scale structural layouts and global high-frequency distributions. It implements a down-sampling via three sequential strides, significantly compressing spatial size to extract compact high-level representations.
Let $G_{M}\left(\cdot, \bm{d}_{s}, C_{in}, C_{out}\right)$ denote the shared branch architecture, where $\bm{d}_{s}$ denotes the sequence of down-sampling strides, $C_{in}$ denotes the number of intermediate channels, and $C_{out}$ denotes the output channel dimension. So, the output of three branches can be uniformly expressed respectively as 
$\mathbf{F}_{f} = G_{M}\left(\mathbf{R}_{HF}, \bm{d}_{s}, C_{in}, C_{f} \right)$, $\mathbf{F}_{m} = G_{M}\left(\mathbf{R}_{HF}, \bm{d}_{s}, C_{in}, C_{m} \right)$, and $\mathbf{F}_{c} = G_{M}\left(\mathbf{R}_{HF}, \bm{d}_{s}, C_{in}, C_{c} \right)$, where $C_{f}$, $C_{m}$ and $C_{c}$ denote the channel dimension of $\mathbf{F}_{f}$, $\mathbf{F}_{m}$ and $\mathbf{F}_{c}$, respectively. 

Let $G_{P}\left( \cdot, \left(h,w\right)\right)$ denote the adaptive average pooling operator that resizes input features to spatial dimensions $(h,w)$. The output of spatial dimensions related to those branches are unified to the target size $\left(H_{t},W_{t}\right)$, namely $\widetilde{\mathbf{F}}_{f}=G_{P}\left( \mathbf{F}_{f}, \left(H_{t},W_{t}\right)\right)$, $\widetilde{\mathbf{F}}_{m}=G_{P}\left( \mathbf{F}_{m}, \left(H_{t},W_{t}\right)\right)$, and $\widetilde{\mathbf{F}}_{c}=G_{P}\left( \mathbf{F}_{c}, \left(H_{t},W_{t}\right)\right)$. The aligned multi-scale features are then concatenated along the channel dimension to form a unified representation, namely $\mathbf{F}_{Cat}=[\widetilde{\mathbf{F}}_{f}, \widetilde{\mathbf{F}}_{m}, \widetilde{\mathbf{F}}_{c}]\in \mathbb{R}^{B \times C_{t} \times H_{t} \times W_{t}}$ with $C_{t}=C_{f}+C_{m}+C_{c}$. 
As to \textit{Branch-fusion} layer, a $1\times 1$ convolutional layer with batch normalization and ReLU activation is applied to fuse the concatenated features and reduce the channel dimension. The finally fused features can be expressed as $\mathbf{F}_{BF} = \sigma_{RL}\left(\sigma_{BN}\left(Conv_{(1,1,0)}\left(\mathbf{F}_{Cat} \right) \right) \right)$, where $\sigma_{RL}\left(\cdot \right)$ denotes ReLU activation operator, and $\sigma_{BN}(\cdot)$ denotes batch normalization operator. 

$\bullet$ \textbf{\textit{Channel Attention} Module}: To adaptively emphasize informative HF channels, we develop a lightweight \textit{Channel Attention} module after cross-scale feature fusion. As to $\mathbf{F}_{BF}$, we generate a set per-channel weights $M_{c} \in \mathbb{R}^{B \times C_{t} \times 1 \times 1}$ and recalibrate the feature maps as follows. First, both global average pooling and global max pooling are applied along the spatial dimensions to aggregate channel-wise statistics, i.e., $\mathbf{F}_{Avg} = \frac{1}{H_{t}W_{t}} \sum_{i=1}^{H_{t}}\sum_{j=1}^{W_{t}}\mathbf{F}_{BF}\left(:,:,i,j \right) \in \mathbb{R}^{B\times C_{t}}$ and $\mathbf{F}_{Max} = \operatorname*{max}_{i \in \left[1, H_{t}\right], j\in \left[1,W_{t} \right]  } \mathbf{F}_{BF}\left(:,:,i,j\right) \in \mathbb{R}^{B\times C_{t}}$. 
These feature maps are then passed through a shared multi-layer perceptron (MLP) operator to generate the channel attention, and the finally recalibrated feature $\mathbf{F}_{Attn}$ is obtained via channel-wise multiplication, which can be expressed as 
\begin{equation}
\mathbf{F}_{Attn} = \sigma_{SM}\left(G_{MLP}\left(\mathbf{Z}_{Avg} \right)+G_{MLP}\left(\mathbf{Z}_{Max} \right)   \right) \otimes \mathbf{F}_{BF},
\setcounter{equation}{10}
\label{eq10}
\end{equation}
where $\otimes$ denotes element-wise multiplication, $\sigma_{SM}\left(\cdot \right)$ denotes the Sigmoid activation operator, and the $G_{MLP}(\cdot)$ consists of two $1\times 1$ convolutional layers and ReLU activation operator, i.e., $G_{MLP}(\cdot)=Conv_{(1,1,0)}\left(\sigma_{RL}\left(Conv_{(1,1,0)}\left(\cdot \right) \right) \right)$.

To stabilize gradient flow, a global residual connection injects the original HR image as an HF prior. The projected prior is then fused into the HF feature via a learnable scaling factor $\psi_{HF}$, which can be expressed as
\begin{equation}
\mathbf{Z}_{HF} = \mathbf{F}_{Attn} + \psi_{HF}\cdot \mathbf{S}_{Proj},
\setcounter{equation}{11}
\label{eq11}
\end{equation}
where $\mathbf{S}_{Proj}=\sigma_{BN}\left(Conv_{(1,1,0)}\left(G_{P}\left(\mathbf{S};\left(H_{t}, W_{t} \right) \right) \right) \right)$, and $\psi_{HF}$ denotes a learnable parameter initialized to 0.1.
%which adaptively balances the contributions of HF details and LF structures. 

\subsection{SE-B Design for LF Feature Extraction}

As shown in Fig. 3, SE-B extracts LF semantic features via the Swin Transformer, hierarchically aggregating structurally consistent features from LR inputs by synergizing convolutional down-sampling with window-based self-attention.

As to the \textit{Pre-embedding} module, the raw $\mathbf{S}_{\downarrow}$ is first projected into a high-dimensional embedding space to facilitate subsequent feature transformations, which can be expressed as $\mathbf{F}_{Pemb} = Conv_{(1,1,0)}\left(\sigma_{RL}\left(\sigma_{BN}\left( Conv_{(3,1,1)}\left(\mathbf{S}_{\downarrow} \right)\right) \right) \right)$.
Then, the following employs two cascaded stages, and each of the two cascaded stages comprises a \textit{Down-sampling$\And$Dynamic-padding} module and a \textit{Swin Transformer Block} module. Specifically, depth-wise separable convolution down-samples while expanding channels via $1\times1$ convolution; Dynamic-padding accommodates arbitrary input resolutions before the \textit{Swin Transformer Block} processing. 
Taking the first stage as an example, its output can be expressed as 
\begin{equation}
\mathbf{F}_{Swin}^{(1)} = G_{S_{2}}\left(G_{S_{1}}\left(G_{Win}\left(G_{Pad}\left(\mathbf{F}_{Ds};w \right) \right) \right) \right),
\setcounter{equation}{12}
\label{eq12}
\end{equation}
where $\mathbf{F}_{Ds} = Conv_{(1,1,0)}\left(DConv_{(4,2,1)}\left(\mathbf{F}_{Pemb} \right) \right)$ denotes the down-sampling operation, $G_{Pad}\left(\cdot;w\right)$ denotes padding operator that the width and height of the input feature are divisible by $w$, $G_{Win}$ denotes the window partition operator that makes padded tensor be non-overlapping windows of size $w\times w$, and the block $\left(G_{S_{1}}, G_{S_{2}} \right)$ applies window-based multi-head self-attention (W-MSA) and shifted-window MSA (SW-MSA), 
which computes self-attention within non-overlapping local windows for efficiency while shifts these windows for capturing long-range dependencies [34].

\begin{figure}[t]
\centering
\resizebox{0.42\textwidth}{0.06\textheight}{\includegraphics{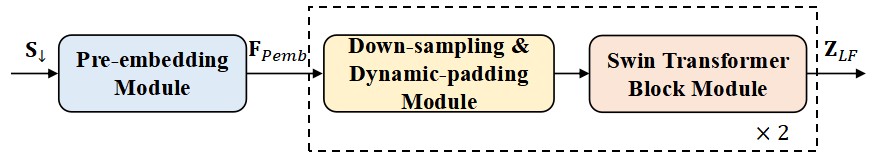}}
\captionsetup{font={footnotesize}}
\caption{Architecture of SE-B for LF feature extraction.}
\label{fig3}
\end{figure}

\subsection{SD Design for Feature Reconstruction}

As depicted in Fig. 4(a), the SD consists of  \textit{Feature Fusion} module and cascaded \textit{Swin Transformer Block} and \textit{SNR Adaptive} modules. 
%\textit{Swin Transformer Block}, and \textit{SNR Adaptive} modules.

$\bullet$ \textbf{\textit{Feature Fusion} Module}:
To reconcile quantized LF structural priors with HF residual details, the proposed gating-inspired \textit{Feature Fusion} module adopts lightweight layers to adaptively harmonize multi-scale features, as detailed in Fig. 4(b).
The quantized LF feature $\widehat{\mathbf{Z}}_{LF}$ and HF feature $\widehat{\mathbf{Z}}_{HF}$ are projected into the \textit{LF-align} and \textit{HF-align} layers, respectively, each implemented as a $1\times1$ convolution with batch normalization and ReLU activation to eliminate channel-dimension discrepancies. Their output features are expressed as
$\mathbf{F}_{LF} = U_{b}(\sigma_{RL}(\sigma_{BN}( Conv_{(1,1,0)}(\widehat{\mathbf{Z}}_{B}))))$ and $\mathbf{F}_{HF} = \sigma_{RL}(\sigma_{BN}(Conv_{(1,1,0)}(\widehat{\mathbf{Z}}_{T} ))) $. 
A lightweight \textit{Gate} layer then processes the concatenated tensor $[\mathbf{F}_{LF},\mathbf{F}_{HF}]$ to produce per-channel modulation weights. It comprises two $3\times3$ convolutions with batch normalization and ReLU, followed by a Sigmoid activation that constrains the weights to $[0,1]$. The gate map is expressed as $\{\mathbf{F}_{GLF}, \mathbf{F}_{GHF}\} = \sigma_{SM}(Conv_{(3,1,1)}(\sigma_{BN}(Conv_{(3,1,1)}[\mathbf{F}_{LF}, \mathbf{F}_{HF}])))$. 
Finally, the aligned features undergo channel-wise adaptive weighting and residual refinement to suppress fusion artifacts, i.e., 
\begin{equation}
\mathbf{Z}_{FF} = \mathbf{F}_{G} + \psi_{FF} \cdot G_{RF}\left(\mathbf{F}_{G} \right),
\setcounter{equation}{13}
\label{eq13}
\end{equation}
where $\mathbf{F}_{G} = \mathbf{F}_{GLF} \otimes  \mathbf{F}_{LF} + \mathbf{F}_{GHF} \otimes  \mathbf{F}_{HF}$, $G_{RF}(\cdot)=\sigma_{BN}(Conv_{(3,1,1)}(\sigma_{R}( \sigma_{BN}(Conv_{(3,1,1)}(\cdot)))))$ denotes the refinement convolutional block defined in \textit{Refine} layer, and $\psi_{FF}$ denotes a learnable scalar initialized to 0.1.

\begin{figure}
\centering
\subfigure[]{
  \label{fig:subfig:a}
   \includegraphics[width = 0.395\textwidth]{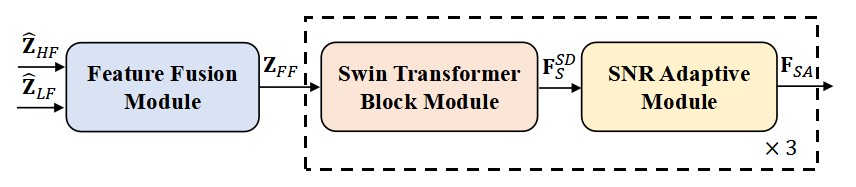}}
\hspace{1in}
\subfigure []{
 \label{fig:subfig:b}
 \includegraphics[width = 0.36\textwidth]{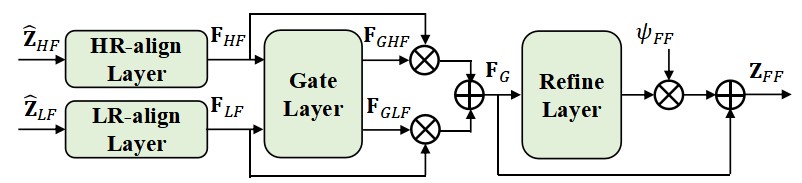}}
 \hspace{1in}
\subfigure []{
 \label{fig:subfig:b}
 \includegraphics[width = 0.32\textwidth]{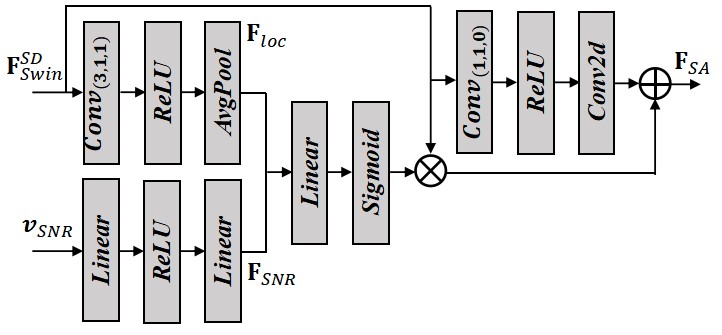}}
\captionsetup{font={footnotesize}}
\caption{
Architecture of SD for feature reconstruction. (a) Overall structure, where HF and LF features are fused by \textit{Feature Fusion} module followed by cascaded \textit{Swin Transformer Block} and \textit{SNR Adaptive} modules. (b) Structure of \textit{Feature Fusion} module. (c) Structure of \textit{SNR Adaptive} module.
} 
\label{fig:subfig}
\end{figure}

$\bullet$ \textbf{\textit{SNR Adaptive} Module}: The detailed structure of \textit{SNR Adaptive} module is shown in Fig. 4(c). 
Prior to entering the \textit{SNR Adaptive} module, the features are processed by \textit{Swin Transformer Block} module,
i.e., $\mathbf{F}_{Swin}^{SD} = G_{S_{2}}\left(G_{S_{1}}\left(G_{Win}\left(G_{Pad}\left( \mathbf{Z}_{FF}; w\right)\right) \right)\right)$.
Let $\bm{v}_{SNR} \in \mathbb{R}^{B}$ denote the batch-wise SNR value. The scalar SNR value is projected into a high-dimensional semantic space compatible with the feature channels, i.e., $\mathbf{Z}_{SNR} = \sigma_{LN}\left(\sigma_{RL}\left(\sigma_{LN}\left( \bm{v}_{SNR}\right) \right) \right)$,
where $\sigma_{LN}\left(\cdot \right)$ denotes linear operator.
A shallow convolutional branch extracts local descriptors, which are aggregated into global satistics via adaptive average pooling as $\mathbf{F}_{loc} = G_{P}(\sigma_{R}( Conv_{(3,1,1)}(\mathbf{F}_{S}^{SD})); (1,1) )$.
The local feature context and SNR embedding are concatenated and fed into a gating network to predict channel-wise scaling factors, i.e., $\mathbf{F}_{LS}=\sigma_{S}(\sigma_{LN}([\mathbf{F}_{loc},\mathbf{F}_{SNR}]))$. The output features are then adaptively recalibrated as 
\begin{equation}
\mathbf{F}_{SA} = \mathbf{F}_{S}^{SD}  \otimes \mathbf{F}_{LS} + Conv_{(1,1,0)}(\sigma_{LR}(Conv_{(1,1,0)}(\mathbf{F}_{S}^{SD}))). 
\setcounter{equation}{14}
\label{eq14}
\end{equation}
Finally, the reconstructed image $\widehat{\mathbf{S}}$ is obtained through three cascaded processing stages.

\subsection{Training Strategy Over AFDM-based Satellite Channels}

The DVQ-SDSC framework must be trained under practical satellite channels and additive-noise conditions. In particular, a single bit error in the demodulated symbol $\widehat{\bm{x}}_{d}$ induces an index flip in $\widehat{\bm{I}}_{s}$, and then this error propagates through the embedding lookup, generating a semantically incorrect codeword.
Moreover, the total AFDM frame length is shared among pilot, guard, and data symbols. Increasing the pilot density improves channel-estimation accuracy at the cost of reduced bandwidth for semantic transmission, thereby intensifying the already severe bandwidth constraint. 

Recent studies formulate AFDM channel estimation as sparse signal recovery due to the inherent sparsity of $\mathbf{H}_{e}$, where the pilot arrangement and sparse measurement matrix design are key considerations.
The DAFT-domain transmit vector $\boldsymbol{x}\in\mathbb{C}^{N}$ is assembled by concatenating pilot, guard, and data symbols. Let the leading $N_{p}$ entries, $k\in\{0,\ldots,N_{p}-1\}$, hold the pilot symbols $x_{p}[k]$ and the next $N_{g}$ entries, $k\in\{N_{p},\ldots,N_{p}+N_{g}-1\}$, and the last $N_{g}$ entries, $k\in\{N_{p}+N_{g}+N_{d},\ldots,N-1\}$, serve as guard symbols that isolate the channel-induced pilot spread from the adjacent data block. 
The $N_d$ data symbols $x_d[k]$ occupy the central $N_d$ entries, i.e., $k\in\{N_p+N_g,\ldots,N-N_g-1\}$.
The AFDM receiver extracts a contiguous $N$-point DAFT-domain observation window for sparse channel estimation, starting at index $\widetilde{m}_s=k_{\max}+N_v-N_g$, where $k_{\max}$ and $N_v$ denote the maximum normalized Doppler shift and the number of Doppler carriers, respectively.
Given the guard length $N_g=(l_{\max}+1)(2k_{\max}+2N_v-1)$ with $l_{\max}$ being the maximum normalized delay, the total number of observed symbols is $N_T=N_p+N_g$.
Extracted via modulo-$N$ indexing, the window has index set $\bm{I}_{T}=\{(\widetilde{m}_{s}+i)\bmod N\mid i=0,\ldots,N_{T}-1\}$. From \eqref{eq6}, the received pilot signal $\bm{y}_{T}\in\mathbb{C}^{N_{T}}$ is rewritten as

\begin{algorithm}[t]
\caption{Training Strategy of DVQ-SDSC}
\label{alg:training}
\footnotesize
\begin{algorithmic}[1]

\REQUIRE Image dataset $\mathcal{D}$; AFDM channel $\mathbf{H}_e$ with pilot arrangement; noise power $\sigma^2$;
        SE-T $f_{SE\text{-}T}(\cdot,\alpha_T)$, SE-B $f_{SE\text{-}B}(\cdot,\alpha_B)$;
        trainable codebooks $\mathbf{C}_{HF}$, $\mathbf{C}_{LF}$; SD model $f_{SD}(\cdot,\beta)$.
\ENSURE  Optimized parameters $\{\alpha_T^*,\alpha_B^*,\beta^*\}$ and codebooks $\{\mathbf{C}_{HF}^*,\mathbf{C}_{LF}^*\}$.

% ==================== Stage 1 ====================
\STATE \textbf{Stage 1 (Codec \& Codebook Pretraining):}
\STATE Initialize $\alpha_T,\alpha_B,\beta,\mathbf{C}_{HF},\mathbf{C}_{LF}$.
\WHILE{not converged}
    \STATE Sample batch $\{S,S_\downarrow\}$ from $\mathcal{D}$.
    \STATE Extract $\mathbf{Z}_{HF}\!=\!f_{SE\text{-}T}(S,S_\downarrow,\alpha_T)$, $\mathbf{Z}_{LF}\!=\!f_{SE\text{-}B}(S_\downarrow,\alpha_B)$ by \eqref{eq1}.
    \STATE Quantize features to indices $\bm{I}^s\!=\!\{\bm{I}_{HF}^s,\bm{I}_{LF}^s\}$ via VQ by \eqref{eq2}.
    \STATE Modulate $\bm{I}^s\!\to\!\bm{x}_d$ and transmit over AFDM (ideal channel). %and receive $\bm{y}_d$.
    \STATE Reconstruct $\widehat{\mathbf{S}}=f_{SD}(\widehat{\mathbf{Z}}_{HF},\widehat{\mathbf{Z}}_{LF},\beta)$ by \eqref{eq7}.
    \STATE Compute $\mathcal{L}_{rec}=d(S,\widehat{\mathbf{S}})$ by \eqref{eq8}.
    \STATE Update $\alpha_T,\alpha_B,\beta,\mathbf{C}_{HF},\mathbf{C}_{LF}\leftarrow\min\mathcal{L}_{rec}$.
\ENDWHILE
\STATE Set $\alpha_T^*\!\leftarrow\!\alpha_T,\;\alpha_B^*\!\leftarrow\!\alpha_B,\;\mathbf{C}_{HF}^*\!\leftarrow\!\mathbf{C}_{HF},\;\mathbf{C}_{LF}^*\!\leftarrow\!\mathbf{C}_{LF}$.

% ==================== Stage 2 ====================
\STATE \textbf{Stage 2 (SD Refinement under Satellite Channel):}
\STATE Freeze $\alpha_T^*,\alpha_B^*,\mathbf{C}_{HF}^*,\mathbf{C}_{LF}^*$ (encoder \& codebooks fixed).
\WHILE{not converged}
    \STATE Modulate $\mathbf{I}^s\!\to\!\bm{x}_d$, transmit over $\mathbf{H}_e$ with noise $\sigma^2$.
    \STATE Extract observation window $\bm{y}_T$ from received DAFT-domain signal.
    \STATE Estimate $\widehat{\mathbf{H}}_e$ via OMP sparse recovery by \eqref{eq18}.
    \STATE Equalize and demodulate to recover $\widehat{\mathbf{I}}^s$.
    \STATE Reconstruct $\widehat{\mathbf{S}}=f_{SD}(\widehat{\mathbf{Z}}_{HF},\widehat{\mathbf{Z}}_{LF},\beta)$ by \eqref{eq7}.
    \STATE Update $\beta\leftarrow\min\mathcal{L}_{rec}=d(S,\widehat{\mathbf{S}})$.
\ENDWHILE
\STATE Set $\beta^*\leftarrow\beta$.
\end{algorithmic}
\textbf{Output: $\{\alpha_T^*,\alpha_B^*,\beta^*\}$ and $\{\mathbf{C}_{HF}^*,\mathbf{C}_{LF}^*\}$.}
\end{algorithm}

\begin{equation}
\bm{y}_{T} = \bm{\Phi}\left(\bm{g}\right)\bm{h} + \bm{w}_{T},
\setcounter{equation}{15}
\label{eq15}
\end{equation}
where $\bm{h}=[h_{0},...,h_{L_{p}-1}]^{T} \in \mathbb{C}^{L_{p}}$, 
$\bm{w}_{T} \in \mathbb{C}^{N_{T}}$ denotes the measurement noise, and $\bm{\Phi}\left(\bm{g}\right) = [\bm{\phi}(g_{0}),...,\bm{\phi}(g_{i}),...,\bm{\phi}(g_{L_{p}-1})]\in \mathbb{C}^{N_{T}\times L_{p}} $ denotes the measurement matrix. The $\widetilde{m}$th entry of $\bm{\phi}\left(g_{i}\right)$ is calculated as 
\begin{align}
\bm{\phi}_{\widetilde{m}}\left(g_{i}\right) = \frac{1}{N}\sum_{m\in \{0,...,N_{p}-1\}}\bm{x}_{m} e^{j2\pi \left[c_{1}l_{i}^{2} - \frac{ml_{i} }{N } + c_{2}\left(m^{2} - \widetilde{m}^{2} \right)      \right]}  \nonumber\\
\times \frac{e^{-j2\pi \left[\widetilde{m} - m + 2Nc_{1}l_{i} + \bar{\nu}_{i}+\iota_{i}\right]} -1  }{e^{-j\frac{2\pi}{N} \left[\widetilde{m} - m + 2Nc_{1}l_{i} + \bar{\nu}_{i}+\iota_{i}\right]} -1   },
\setcounter{equation}{15}
\label{eq16}
\end{align}
Since the measurement matrix $\bm{\Phi}(\bm{g})$ depends on the unknown channel information $\{l_i,\bar{\nu}_i,\iota_i\}_{i=0}^{L_p-1}$, direct channel-parameter estimation from \eqref{eq15} is infeasible. To overcome this, we transform channel estimation into a sparse signal recovery problem over a virtual sampling grid. Specifically, we sample the delay and Doppler dimensions over $[0,l_{\max}]$ and $[-\nu_{\max},\nu_{\max}]$ with resolutions $r_{\tau}=l_{\max}/(N_{\tau}-1)$ and $r_{\nu}=(2\nu_{\max}+2)/(N_{\nu}-1)$, yielding the delay and Doppler vectors $\widetilde{\bm{l}}=[\widetilde{l}_0,\ldots,\widetilde{l}_{N_{\tau}-1}]^T$ and $\widetilde{\bm{k}}=[\widetilde{k}_0,\ldots,\widetilde{k}_{N_{\nu}-1}]^T$, where $\widetilde{l}_a=ar_{\tau}$ and $\widetilde{k}_b=br_{\nu}-k_{\max}-1$. We then define the virtual sampling grid as $\widetilde{S}_i=\{\widetilde{l}_i,\widetilde{k}_i\}_{i}^{N_s}$ with $N_s=N_{\tau}N_{\nu}$. The AFDM channel estimation model is thus established as
%So, the measurement matrix $\bm{\Phi}\left(\bm{g}\right)$ is characterized based on unkown channel information $\left\{   l_{i},\bar{\nu}_{i},\iota_{i} \right\}_{i=0}^{L_{p}-1}$. Therefore, it is infeasible to directly estimate the channel parameters based on (18).
%To overcome this challenge, we transform the channel estimation into a sparse signal recovery problem by establishing a virtual sampling grid. 
%Specifically, we first perform the virtual sampling along the delay and Doppler dimension in the range of $\left[0,l_{max}\right]$ and $\left[-\nu_{max}, \nu_{max} \right]$ with delay resolution $r_{\tau} = \frac{l_{max} }{N_{\tau} -1}$ and Doppler resolution $r_{\nu}=\frac{2\nu_{max} + 2}{N_{\nu} -1}$. Then, the virtual sampling delay and Doppler vectors can be given $\widetilde{\bm{l}} = \left[\widetilde{l}_{0},...,\widetilde{l}_{N_{\tau} -1}   \right]^{T}$ and $\widetilde{\bm{k}} = \left[\widetilde{k}_{0},..., \widetilde{k}_{N_{\nu} -1}\right]^{T} $ with $\widetilde{l}_{a}=ar_{\tau},a=0,...,N_{\tau} -1$ and $\widetilde{k}_{b}=br_{\nu} - k_{max} -1,b=0,...,N_{\nu} -1$. Furthermore, we define the virtual sampling grid as $\widetilde{S}_{i}=\left\{\widetilde{l}_{i},\widetilde{k}_{i}  \right\}_{i}^{N_{s}}$ with $N_{s}=N_{\tau}N_{\nu}$. The AFDM channel estimation model can be established as
\begin{equation}
\bm{y}_{T} = \bm{\Phi}\left(\bm{\widetilde{g}}\right)\widetilde{\bm{h}} + \bm{w}_{T},
\setcounter{equation}{17}
\label{eq17}
\end{equation}
where $\bm{\Phi}(\widetilde{\bm{g}})=[\bm{\phi}(\widetilde{g}_0),\ldots,\bm{\phi}(\widetilde{g}_{N_s-1})]\in\mathbb{C}^{N_T\times N_s}$ denotes the measurement matrix over the virtual sampling grid $\widetilde{\bm{S}}$, and $\widetilde{\bm{h}}=[\widetilde{h}_0,\ldots,\widetilde{h}_{N_s-1}]^T\in\mathbb{C}^{N_s}$ the channel vector to be estimated. Thus, the sparse channel estimation is formulated as
%where $\bm{\Phi}\left(\bm{\widetilde{g}}\right)=\left[\bm{\phi}\left(\widetilde{g}_{0}\right),...,\bm{\phi}\left(\widetilde{g}_{i}\right),...,\bm{\phi}\left(\widetilde{g}_{N_{s}-1}\right)  \right]\in \mathbb{C}^{N_{T}\times N_{s}}$ denotes the measurement matrix based on virtual sampling grid $\bm{\widetilde{S}}$, and $\widetilde{\bm{h}} = \left[\widetilde{h}_{0},...,\widetilde{h}_{i},...,\widetilde{h}_{N_{s}-1}   \right]^{T} \in \mathbb{C}^{N_{s}}$ denotes the channel vector to be estimated. 
%Note that only $L_{p}$ elements are non-zeros.
%So, the sparse channel estimation can be formulated as 
\begin{equation}
\operatorname*{min}_{\widetilde{\bm{h}} \in \mathbb{C}^{N_{s}}  } \| \widetilde{\bm{h}} \|_{0} \hspace{0.3cm}\mathrm{s.t.} \hspace{0.3cm} \| \bm{y}_{T} -  \bm{\Phi}\left(\bm{\widetilde{g}}\right)\widetilde{\bm{h}}\|_{2}\leq \epsilon,
\setcounter{equation}{18}
\label{eq18}
\end{equation}
where $\|\cdot\|_{0}$ denotes zero-norm operation, and $\epsilon$ denotes the noise power bound.

For computational efficiency, we solve \eqref{eq18} via orthogonal matching pursuit (OMP) [35], whose greedy nature is well suited to the sparse effective channel $\mathbf{H}_{e}$, while the codebooks are trained following the procedure in [25]. 
%To fully decouple channel estimation from the semantic codec, the DVQ-SDSC adopts a two-stage training strategy. In Stage 1, the semantic encoder/decoder and the quantized codebooks are jointly optimized without the LEO channel, so that feature extraction and quantization are isolated from channel-induced gradients. In Stage 2, the semantic decoder is retrained under the LEO channel with the encoder and codebooks frozen, allowing the model to adapt to realistic channel impairments while preserving the learned semantic representation. The overall procedure is summarized in \textbf{Algorithm 1}.
To fully decouple channel estimation from the semantic codec, the DVQ-SDSC adopts a two-stage training strategy. In Stage 1, the semantic encoder/decoder and codebooks are jointly optimized without the satellite channel, isolating feature extraction and quantization from channel-induced gradients. In Stage 2, the semantic decoder is retrained under the satellite channel with encoder and codebooks frozen, adapting to realistic channel impairments while preserving learned semantic representation. The overall procedure is summarized in \textbf{Algorithm 1}.

\section{Codebook-Reordering-Aided G-DPCM for Index Compression and Transmission}

The proposed DVQ-SDSC framework establishes a discrete semantic space via a learnable codebook for efficient index-based transmission. 
However, conventional VQ-based methods transmit each index with a fixed bit-length, thereby introducing inherent coding redundancy.
By exploiting the strong spatial correlation among VQ-generated indices arising from locally smooth latent features, we propose a codebook-reordering-aided G-DPCM scheme that encodes predicted residuals instead of absolute index values for further compression.
Specifically, the DPCM-aware PCA method is first introduced to reorder the codebook, aligning its topological structure with the correlation pattern exploited by DPCM, and then the G-DPCM codec is developed to transmit VQ indices over noisy channels with a controllable trade-off between bandwidth efficiency and error resilience.

%The proposed DVQ-SDSC framework establishes a discrete semantic space via a learnable codebook, enabling efficient index-based transmission. Due to the local smoothness of encoded latent features, the VQ process generates a sequence of discrete indices whose spatial adjacency inherently exhibits strong correlation. However, transmitting each index independently with a fixed-length bit disregards this statistical dependency, resulting in substantial redundancy at the transmission side. By exploiting the underlying inter-correlation of index sequence, we develop the group-quantized differential pulse-code modulation (G-DPCM) scheme to encode the predicted residuals rather than absolute index values. Specifically, the DPCM-aware principal component analysis (PCA) method is first introduced to reorder the codebook, which aims to effectively align the topological structure of the discrete index set with the underlying correlation pattern exploited by DPCM, and then the G-DPCM codec is developed to transmit VQ indices over noisy wireless channels while maintaining a predictable trade-off between bandwidth efficiency and error resilience.

\begin{algorithm}[t]
\caption{DPCM-Aware PCA for Codebook Reordering}
\label{alg:dpcm_pca}
\footnotesize
\begin{algorithmic}[1]

\REQUIRE  Original VQ codebook $\mathbf{C}_{HF}=[\boldsymbol{c}_0,\dots,\boldsymbol{c}_{K_T-1}]^T\in\mathbb{R}^{K_T\times N_T}$;
          ring-neighborhood decay parameter $\sigma$.
          %(optional) number $R$ of ring-cut candidates.
%\ENSURE  Optimal permutation $\boldsymbol{\pi}\in\Omega_{K_T}$ and reordered codebook $\mathbf{C}_{HF}^{\boldsymbol{\pi}}$.

% ---------- 1. center the codebook ----------
\STATE Compute the codebook mean $\boldsymbol{\mu}_c=\dfrac{1}{K_T}\sum_{k=0}^{K_T-1}\boldsymbol{c}_k$.
\STATE Center the codewords: $\widetilde{\mathbf{C}}_{HF}=[\tilde{\boldsymbol{c}}_0,\dots,\tilde{\boldsymbol{c}}_{K_T-1}]^T$,
       where $\tilde{\boldsymbol{c}}_k=\boldsymbol{c}_k-\boldsymbol{\mu}_c$.

% ---------- 2. ring-weighted covariance ----------
\STATE Compute the ring-weighted covariance
       $\boldsymbol{\Sigma}_C=\dfrac{1}{K_T-1}\,\widetilde{\mathbf{C}}_{HF}^T\,\mathbf{W}\,\widetilde{\mathbf{C}}_{HF}$
       with $[\mathbf{W}]_{i,j}=\exp\!\big(-d^2(i,j)/(2\sigma^2)\big)$,
       $d(i,j)=\min\{|i-j|,\,K_T-|i-j|\}$.

% ---------- 3. principal direction ----------
\STATE Obtain the leading eigenvector $\boldsymbol{v}_0$ of $\boldsymbol{\Sigma}_C$ (e.g. via power iteration).
\STATE Set $\mathbf{v}=\boldsymbol{v}_0/\|\boldsymbol{v}_0\|$.

% ---------- 4. PCA projection + linear order ----------
\FOR{$k=0$ \TO $K_T-1$}
    \STATE $p_k=\tilde{\boldsymbol{c}}_k^T\mathbf{v}$ \COMMENT{PCA projection score}
\ENDFOR
\STATE $\boldsymbol{\pi}_{\mathrm{lin}}=\operatorname{argsort}(p_0,\dots,p_{K_T-1})$ \COMMENT{ascending order}

% ---------- 5. optimal ring cut ----------
\FOR{each candidate cut $r_p\in\{0,1,\dots,K_T-1\}$}
    \STATE $\boldsymbol{\pi}_{r_p}=\operatorname{roll}(\boldsymbol{\pi}_{\mathrm{lin}},-r_p)$ \COMMENT{left cyclic shift}
    \STATE Evaluate the adjacent-distance cost $\mathcal{J}(r_p)$ by \eqref{eq21}   
\ENDFOR
\STATE $r_p^*=\displaystyle\arg\min_{r_p}\;\mathcal{J}(r_p)$,
       and set the final permutation $\boldsymbol{\pi}=\boldsymbol{\pi}_{r_p^*}=\operatorname{roll}(\boldsymbol{\pi}_{\mathrm{lin}},-r_p^*)$.

% ---------- 6. reordered codebook ----------
\STATE Construct the reordered codebook
       $\mathbf{C}_{HF}^{\boldsymbol{\pi}}=[\boldsymbol{c}_{\boldsymbol{\pi}(0)},\dots,\boldsymbol{c}_{\boldsymbol{\pi}(K_T-1)}]^T$.

%\STATE \textbf{Output:} Optimal permutation $\boldsymbol{\pi}$ and reordered codebook $\mathbf{C}_{HF}^{\boldsymbol{\pi}}$.

\end{algorithmic}
\KwOut{Optimal permutation $\bm{\pi}$ and reordered codebook $\mathbf{C}_{HF}^{\pi}$.}
\end{algorithm}

\subsection{DPCM-Aware PCA Method for Codebook Reordering}
 
Encoded image features generally exhibit strong spatial local correlation. However, standard VQ codebook mapping assigns each feature to its nearest code vector without guaranteeing that similar features obtain neighboring indices. Since the original codebook indices carry no inherent topological meaning, spatially or semantically adjacent regions may map to substantially different entries, leaving the spatial correlation unexploited for compression.
%To formalize this, we introduce the HF feature matrix $\mathbf{Z}_{HF}=[\bm{z}_0,\ldots,\bm{z}_{K_z-1}]^T\in\mathbb{R}^{K_z\times M_z}$, its codebook $\mathbf{C}_{HF}=[\bm{c}_0,\ldots,\bm{c}_{K_T-1}]^T\in\mathbb{R}^{K_T\times N_T}$, and the quantized features $\widehat{\mathbf{Z}}_{HF}=[\widehat{\bm{z}}_0,\ldots,\widehat{\bm{z}}_{K_z-1}]^T$. We then seek a permutation $\bm{\pi}\in\mathbf{\Omega}_{K_z}$ that aligns adjacent indices with nearby codewords while respecting the ring structure (indices 1 and $K_z$ are adjacent), formally stated as
Without loss of generality, we illustrate this using the HF feature matrix $\mathbf{Z}_{HF}=[\bm{z}_0,\ldots,\bm{z}_{K_z-1}]^T\in\mathbb{R}^{K_z\times M_z}$, its corresponding codebook $\mathbf{C}_{HF}=[\bm{c}_0,\ldots,\bm{c}_{K_T-1}]^T\in\mathbb{R}^{K_T\times N_T}$, and the quantized counterpart $\widehat{\mathbf{Z}}_{HF}=[\widehat{\bm{z}}_0,\ldots,\widehat{\bm{z}}_{K_z-1}]^T\in\mathbb{R}^{K_z\times M_z}$. We then seek a permutation $\bm{\pi}\in\mathbf{\Omega}_{K_z}$ that aligns adjacent indices with proximal codewords while preserving the ring structure (where indices 1 and $K_z$ are adjacent), formally stated as
\begin{equation}
\operatorname*{min}_{\bm{\pi} \in \mathbf{\Omega}_{K_{z}}} \sum_{k_{z}=0}^{K_{z}-1}\| \widehat{\bm{z}}_{\bm{\pi}(k_{z})} -\widehat{\bm{z}}_{\bm{\pi}(k_{z}-1)}\|_{2}^{2} \hspace{0.1cm}\mathrm{s.t.} \hspace{0.1cm} \bm{\pi}(0) = \bm{\pi}\left(K_{z}-1 \right),
\setcounter{equation}{19}
\label{eq19}
\end{equation} 
where $\widehat{\bm{z}}_{\bm{\pi}(k_z)}$ denotes the quantized codeword placed at position $k_z$ after permutation $\bm{\pi}$. 

Since the unique vectors in any quantized feature map are exactly the codebook $\mathbf{C}_{HF}$, \eqref{eq19} is equivalent to finding an optimal codebook ordering. Moreover, evaluating \eqref{eq19} directly over data-dependent quantized features would require a different permutation for each input, which is computationally prohibitive. We therefore optimize a single universal permutation by rearranging the codebook vectors $\{\bm{c}_k\}_{k=0}^{K_T-1}$ based on their embedding-space proximity. To avoid an exhaustive permutation search, we impose a ring-adjacency prior on the index space via Gaussian weighting. Let $d(i,j)=\min\{|i-j|,K_z-|i-j|\}$ denote the ring distance between indices $i$ and $j$, and the weight matrix be $[\boldsymbol{W}]_{i,j}=\exp\!\big(-d^2(i,j)/(2\sigma_g^2)\big)$. Denoting the centered codebook by $\widetilde{\mathbf{C}}_{HF}=\mathbf{C}_{HF}-\mathbf{1}_{K_T}\boldsymbol{\mu}_C^T$, where $\boldsymbol{\mu}_C=\frac{1}{K_T}\sum_{k_T=1}^{K_T}\bm{c}_{k_T}$ is the codebook mean, we define the weighted scatter as $\bm{\Sigma}_{C} = \frac{1 }{K_{T}-1 }\widetilde{\mathbf{C}}_{HF}^{T} \mathbf{W} \widetilde{\mathbf{C}}_{HF}$. 
Let $\boldsymbol{v}_0$ be the dominant eigenvector of $\boldsymbol{\Sigma}_C$, obtained via power iteration. Projecting the centered codewords onto $\boldsymbol{v}_0$ yields the scalar scores $p_k=\tilde{\boldsymbol{c}}_k^T\boldsymbol{v}_0$, and the codebook is linearized by sorting these scores in ascending order, i.e.,
\begin{equation}
\boldsymbol{\pi}_{\rm lin}={\rm argsort}\big(\{p_{k}\}_{k=1}^{K_{T}}\big),
\setcounter{equation}{20}
\label{eq20}
\end{equation}
where $\operatorname{argsort}(\cdot)$ returns the permutation that arranges its argument in ascending order.

Since the ordering recovered by \eqref{eq20} is intrinsically circular, a linear layout
is obtained by cutting the ring at a single point. Let
$\boldsymbol{\pi}_{r_{p}}={\rm roll}(\boldsymbol{\pi}_{\rm lin},-r_{p})$ denote
the circularly shifted ordering, where $r_{p}$ is the cut position. The optimal
cut minimizes the total embedded distance between adjacent codewords, i.e.,
\begin{equation}
r_{p}^{*}=\arg\min_{r_{p}}\sum_{k=1}^{K_{T}-1}
\|\tilde{\boldsymbol{c}}_{\pi_{r_{p}}(k)}
-\tilde{\boldsymbol{c}}_{\pi_{r_{p}}(k+1)}\|_{2}^{2},
\setcounter{equation}{21}
\label{eq21}
\end{equation}
where $\tilde{\boldsymbol{c}}_{k}$ is the $k$-th column of
$\widetilde{\mathbf{C}}_{HF}$.

The final codebook permutation is set to $\boldsymbol{\pi}=\boldsymbol{\pi}_{r_p^*}$, obtained by ring-unrolling the circular ordering induced by \eqref{eq20} at the optimal cut point $r_p^*$ determined from \eqref{eq21}. Concretely, we first perform PCA on the Gaussian-weighted covariance matrix $\boldsymbol{\Sigma}_C$, whose Gaussian ring weights $\mathbf{W}$ inherently encode the ring-adjacency prior of DPCM, to extract the dominant eigenvector $\mathbf{v}_0$. 
Sorting all codewords by their projections onto $\mathbf{v}_0$ produces a circular ordering, and then \eqref{eq21} selects the cut point $r_p^*$ minimizing the wrap-around discontinuity, and ring-unrolling at $r_p^*$ yields the linear permutation $\boldsymbol{\pi}_{r_p^*}$. 
The overall procedure is summarized in \textbf{Algorithm 2}. 
The ring-adjacency prior in $\boldsymbol{\Sigma}_C$ ensures that neighboring codewords are placed contiguously in embedding space, which concentrates the prediction residuals.
%and lowers the average bitrate compared with fixed-length coding.

\subsection{G-DPCM Codec for Index Compression and Transmission}

Designed for VQ-index transmission over noisy channels, the proposed G-DPCM codec strikes a controllable balance between bandwidth efficiency and error resilience.  
Departing from conventional entropy-coded DPCM [36] that inherently assumes error-free delivery, the proposed G-DPCM introduces two key innovations: (i) group-wise anchor–residual coding replaces fixed-length index transmission with low-bit-width residual quantization; and (ii) group-based error isolation confines clipping and channel errors within each group, preventing distortion propagation.
The following \textit{Lemma 1} establishes the theoretical foundation for its application.

\textit{Lemma 1 (Residual Concentration and Coding Gain under PCA-aided Codebook Reordering):} For the VQ codebook $\mathbf{C}=\{\bm{c}_k\}_{k=0}^{K_T-1}$ (mean $\bm{\mu}_c$, centered codewords $\tilde{\bm{c}}_k=\bm{c}_k-\bm{\mu}_c$), let $\bm{v}_0$ be the leading eigenvector of the ring-weighted covariance $\bm{\Sigma}_C$ and  $p_k=\tilde{\bm{c}}_k^T\bm{v}_0$ be the scalar projection of each centered codeword. Let $\bm{\pi}$ sort these projections in ascending order,
i.e., $p_{\bm{\pi}(0)}\leq\cdots\leq p_{\bm{\pi}(K_{T}-1)}$, so that
$s_{k}=\bm{\pi}^{-1}(k)$ is the reordered index of $\bm{c}_{k}$. For adjacent latent vectors $\bm{z}_u,\bm{z}_v$ quantized to $\bm{c}_{k_u},\bm{c}_{k_v}$, the folded DPCM residual is defined as $\delta_z=\big((s_{k_v}-s_{k_u})\bmod K_T\big)_{\mathrm{fold}}\in\{-{\lfloor K_T/2\rfloor},\ldots,{\lceil K_T/2\rceil}-1\}$, where $(\cdot)_{\mathrm{fold}}$ symmetrizes the modulo result. Then, the following hold: 
(i) Since adjacent latents in natural images are locally smooth, $\|\bm{z}_{u}-\bm{z}_{v}\|_{2}$ is small with high probability, and so is $\|\bm{c}_{k_{u}}-\bm{c}_{k_{v}}\|_{2}$. 
(ii) Let $b_{a}=\lceil\log_{2}K_{T}\rceil$ be the bit-width of fixed-length absolute index coding. Since $\delta_{z}$ concentrates near zero, a uniform $b_{r}$-bit quantizer with $b_{r}\ll b_{a}$ covers the high-probability residual range without clipping for most index pairs.

\textit{Lemma 1} follows from the variance-maximizing property of the leading eigenvector $\bm{v}_0$. Specifically, codewords close in embedding space yield similar projections $p_k$. Since adjacent latents $\bm{z}_u,\bm{z}_v$ are locally correlated, their nearest codewords lie in nearby regions. Sorting by $p_k$ thus assigns nearby reordered indices to adjacent spatial locations, so that $\delta_z$ concentrates around zero. 
Driven by \textit{Lemma 1}, the overall procedure of G-DPCM is summarized in \textbf{Algorithm 3} and described below.

\textit{1) Encoding Procedure}

Given the reordered index sequence $\bm{s}^r$, the encoder partitions it into $M$ non-overlapping groups $\{\mathcal{G}_m\}_{m=0}^{M-1}$ of $G$ consecutive indices, i.e., $\mathcal{G}_m=\{s^r_{mG},\ldots,s^r_{mG+G-1}\}$. The first index $s^r_{mG}$ of each group serves as an anchor, coded with $b_a$ bits, while the remaining indices are differentially encoded using the raw DPCM residual quantized with $b_r$ bits as
\begin{equation}
\Delta_{n}=s^{r}_{mG+n}-s^{r}_{mG+n-1},\quad n=1,\ldots,G-1.
\setcounter{equation}{22}
\label{eq22}
\end{equation}
To keep the residual within a symmetric range around zero, we apply
modulo-folding as follows
\begin{equation}
\widetilde{\Delta}_{n}=\big(\Delta_{n}\bmod K_{T}\big)_{\mathrm{fold}}
\in\Big\{-\Big\lfloor\tfrac{K_{T}}{2}\Big\rfloor,\ldots,
\Big\lceil\tfrac{K_{T}}{2}\Big\rceil-1\Big\},
\setcounter{equation}{23}
\label{eq23}
\end{equation}
where $x_{\mathrm{fold}}=x$ if $x\leq\lfloor(K_{T}-1)/2\rfloor$ and
$x_{\mathrm{fold}}=x-K_{T}$ otherwise. 

By \textit{Lemma 1(i)}, the folded residual is deterministically bounded by $\widetilde{\Delta}_n\leq B_{\max}$. Under the design condition $B_{\max}\ll K_T/2$, no effective modulo wrap-around occurs for spatially smooth features, and the symmetric interval fully contains the residual dynamic range. The folded residual is then mapped to a $b_r$-bit unsigned integer via an offset-clip operation, i.e.,
\begin{equation}
\widehat{\Delta}_{n}=\mathrm{clip}( \widetilde{\Delta}_{n} + 2^{b_{r}-1},0,2^{b_{r}}-1),
\setcounter{equation}{24}
\label{eq24}
\end{equation}
which denotes a bijection from integers in $[-2^{b_{r}-1},2^{b_{r}-1}-1]$ to $\{0,1,...,2^{b_{r}}-1\}$ via an offset of $2^{b_{r}-1}$. 

\begin{algorithm}[t]
\caption{G-DPCM codec for Index Compression and Transmission}
\label{alg:gdpcm}
\footnotesize
\begin{algorithmic}[1]
\REQUIRE Reordered index sequence $\bm{s}^{r}\in\{0,\ldots,K_T-1\}^{L_s}$ from \textbf{Algorithm 2};
         codebook size $K_T$; group size $G$; anchor bit-width $b_a=\lceil\log_2 K_T\rceil$;
         residual bit-width $b_r$ satisfying $2^{b_r-1}\ge B_{\max}$.
\STATE \textbf{Notation:} $\mathrm{bin}(x,L)$ maps $x\in\{0,\ldots,2^{L}-1\}$ to an
$L$-bit unsigned string, $\mathrm{bin2int}(\cdot)$ is its inverse; $a\bmod K_{T}$
returns the \emph{non-negative} remainder in $\{0,\ldots,K_{T}-1\}$;
$\parallel$ denotes bit-stream concatenation.
%\STATE \textbf{Output:} reconstructed codeword indices $\hat{k}_l=\pi(\hat{s}^r_l)$, $l=0,\ldots,L_s-1$.
% -------- Encoding --------
\STATE \textbf{Encoding Procedure:} 
\STATE $\mathcal{B}\leftarrow\emptyset$; $M\leftarrow\lceil L_s/G\rceil$.
\FOR{$m=0$ \TO $M-1$}
  \STATE $G_m\leftarrow\min\{G,\,L_s-mG\}$;
         $\mathcal{B}\leftarrow\mathcal{B}\parallel\mathrm{bin}(s^r_{mG},\,b_a)$. \label{alg:anchor}
  \FOR{$n=1$ \TO $G_m-1$}
    \STATE $\Delta_n\leftarrow(s^r_{mG+n}-s^r_{mG+n-1})\bmod K_T$.
    \STATE map $\Delta_n$ to signed residual $\widetilde{\Delta}_n\in[-B_{\max},B_{\max}]$ 
    %(centered at $K_T/2$).
    \STATE $\widehat{\Delta}_n\leftarrow\mathrm{clip}(\widetilde{\Delta}_n+2^{b_r-1},\,0,\,2^{b_r}-1)$;
           $\mathcal{B}\leftarrow\mathcal{B}\parallel\mathrm{bin}(\widehat{\Delta}_n,\,b_r)$.
  \ENDFOR
\ENDFOR
% -------- Decoding --------
\STATE \textbf{Decoding Procedure:} 
\STATE $\hat{\bm{s}}^r\leftarrow\mathbf{0}_{L_s}$; $\mathrm{idx}\leftarrow0$.
\FOR{$m=0$ \TO $M-1$}
  \STATE $G_m\leftarrow\min\{G,\,L_s-mG\}$;
         $\hat{s}^r_{mG}\leftarrow\mathrm{bin2int}\big(\mathcal{B}[\mathrm{idx}:\mathrm{idx}+b_a-1]\big)$;
         $\mathrm{idx}\leftarrow\mathrm{idx}+b_a$.
  \FOR{$n=1$ \TO $G_m-1$}
    \STATE $\widetilde{\Delta}_n\leftarrow\mathrm{bin2int}\big(\mathcal{B}[\mathrm{idx}:\mathrm{idx}+b_r-1]\big)-2^{b_r-1}$;
           $\mathrm{idx}\leftarrow\mathrm{idx}+b_r$.
    \STATE $\hat{s}^r_{mG+n}\leftarrow(\hat{s}^r_{mG+n-1}+\widetilde{\Delta}_n)\bmod K_T$.
  \ENDFOR
\ENDFOR
\end{algorithmic}
\KwOut{Reconstructed codeword indices $\hat{k}_{l}\leftarrow\pi\bigl(\hat{s}^{r}_{l}\bigr)$, $l=0,\ldots,L_{s}-1$.}
\end{algorithm}

\textit{2) Decoding Procedure}

At the receiver, the index sequence is recovered group by group. The anchor of each group is decoded from its $b_a$-bit representation, and each received residual is mapped back to the signed folded range by inverting the offset, i.e., $\widetilde{\Delta}_{n}=\widehat{\Delta}_{n}-2^{b_{r}-1}$. 
The remaining indices then follow from the recursion
\begin{equation}
s^{r}_{mG+n}=\big(s^{r}_{mG+n-1}+\widetilde{\Delta}_{n}\big)\bmod K_{T},
\quad n=1,\ldots,G-1,
\setcounter{equation}{25}
\label{eq25}
\end{equation}
and the corresponding codeword is retrieved as $\bm{c}_{\pi(s^{r}_{mG+n})}$. 
Since each group restarts from its anchor, residual errors propagate only within that group and cannot cross its boundaries. 
This group-wise error isolation gives G-DPCM a key advantage over wireless satellite channels.

Without G-DPCM, each of the $L_s$ indices requires $b_a$ bits, i.e., $L_s b_a$ bits in total. With G-DPCM, the sequence is partitioned into $M=\lceil L_s/G\rceil$ groups, each sending one $b_a$-bit anchor followed by $G-1$ residuals of $b_r$ bits, so that $B_{G\text{-}DPCM}=\sum_{m=0}^{M-1}[b_{a}+(G_{m}-1)b_{r}]$, where
$G_{m}=\min\{G,L_{s}-mG\}$ accounts for a possibly truncated last group. For $L_s\gg G$, it reduces to $B_{\text{G-DPCM}}\approx M\,[b_a+(G-1)b_r]$, and the compression ratio is $\rho_{G}=\frac{L_{s} b_{a}}{B_{G\text{-}DPCM}}
\approx\frac{G b_{a}}{b_{a}+(G-1)b_{r}}$.

\section{Experiments and Results Discussion}

This section comprehensively evaluates the proposed DVQ-SDSC framework for high-resolution RIS transmission over AFDM satellite channels, along with the G-DPCM codec in terms of bandwidth efficiency and error resilience.

\begin{figure*}[t] %
\centering
\subfigure[]
{
\begin{minipage}[c]{5.5cm}
    \centering
    \includegraphics[width=1.1\linewidth]{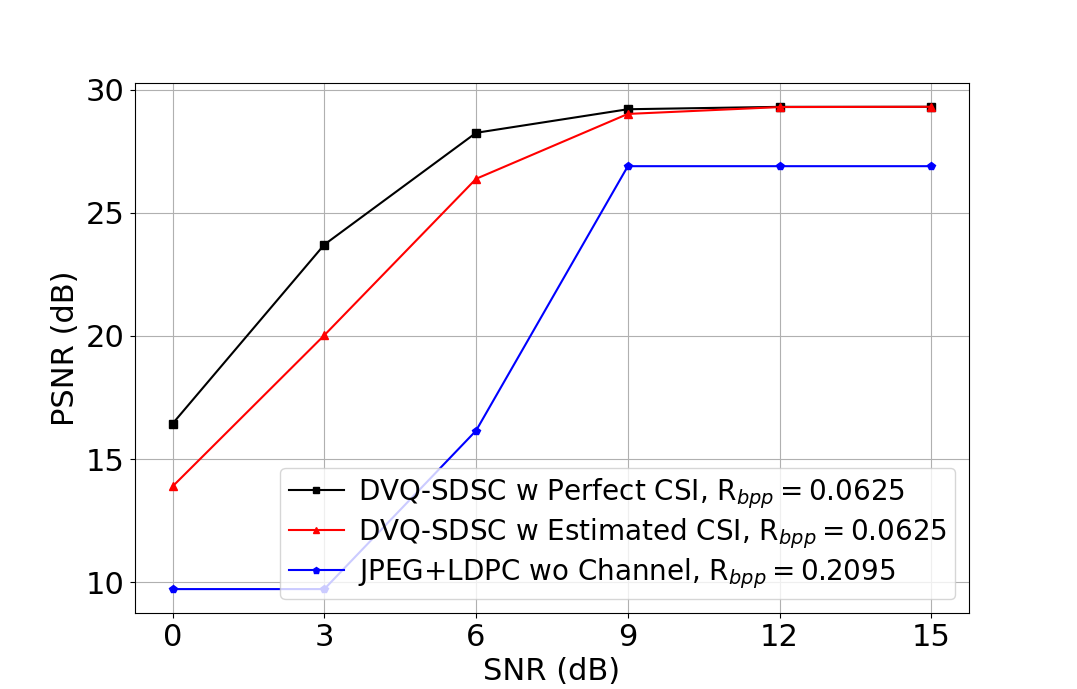}
\end{minipage}
}
\subfigure[]
{
    \begin{minipage}[c]{5.5cm}
    \centering
    \includegraphics[width=1.1\linewidth]{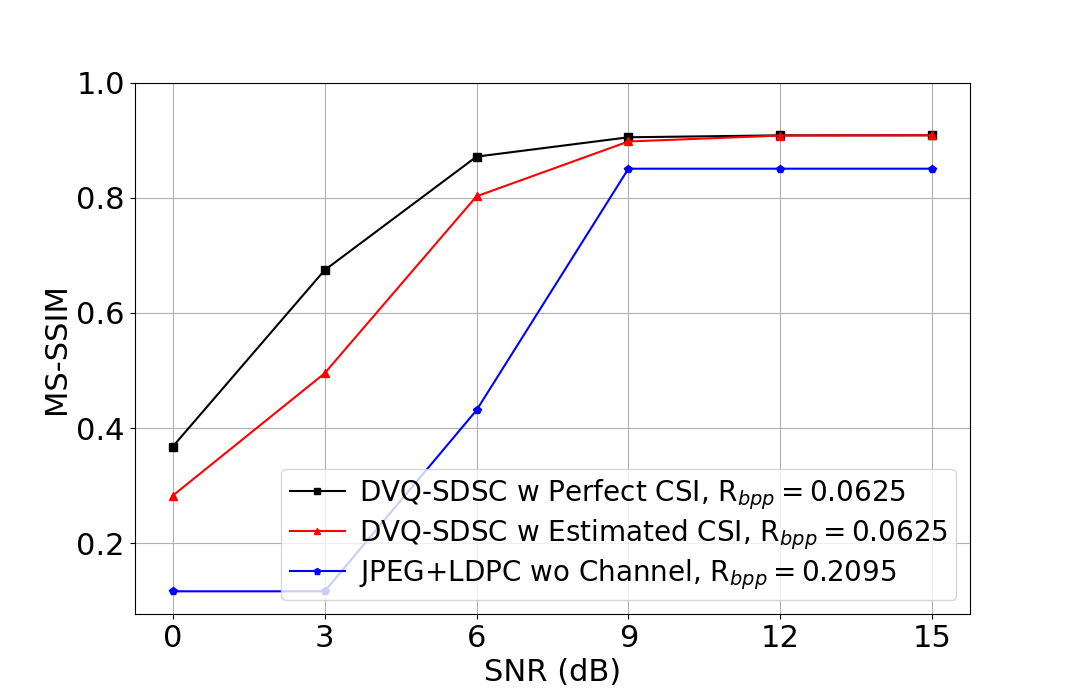}
    \end{minipage}
}
\subfigure[]
{
    \begin{minipage}[c]{5.5cm}
    \centering
    \includegraphics[width=1.1\linewidth]{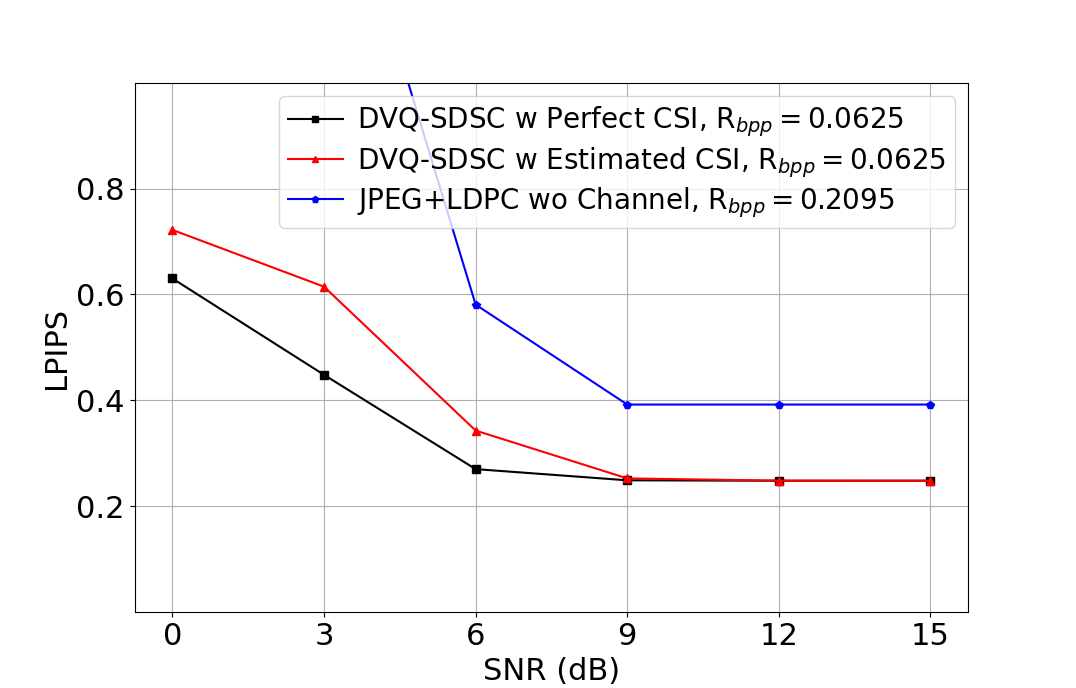}
\end{minipage}
}
\subfigure[]
{
    \begin{minipage}[c]{5.5cm}
    \centering
    \includegraphics[width=1.1\linewidth]{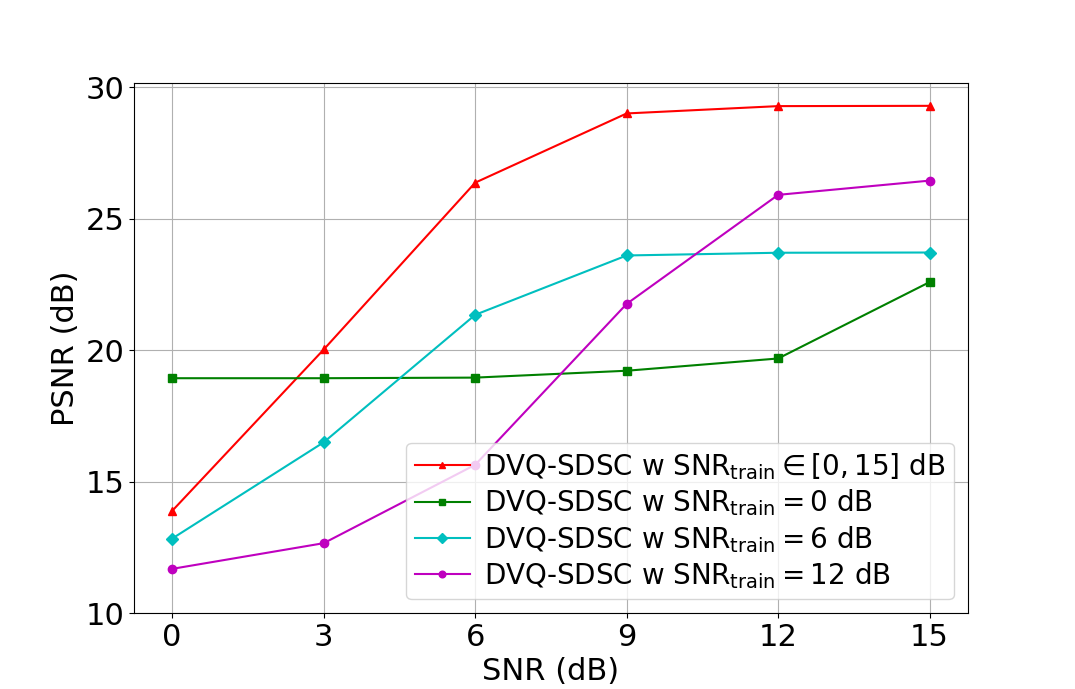}
\end{minipage}
}
\subfigure[]
{
    \begin{minipage}[c]{5.5cm}
    \centering
    \includegraphics[width=1.1\linewidth]{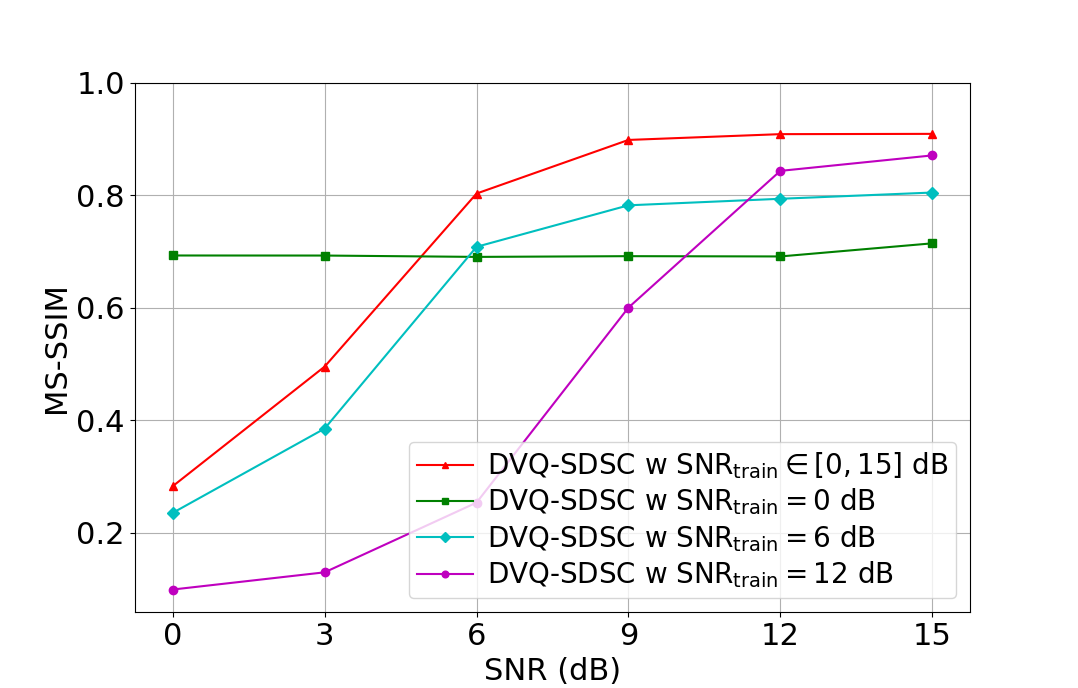}
\end{minipage}
}
\subfigure[]
{
    \begin{minipage}[c]{5.5cm}
    \centering
    \includegraphics[width=1.1\linewidth]{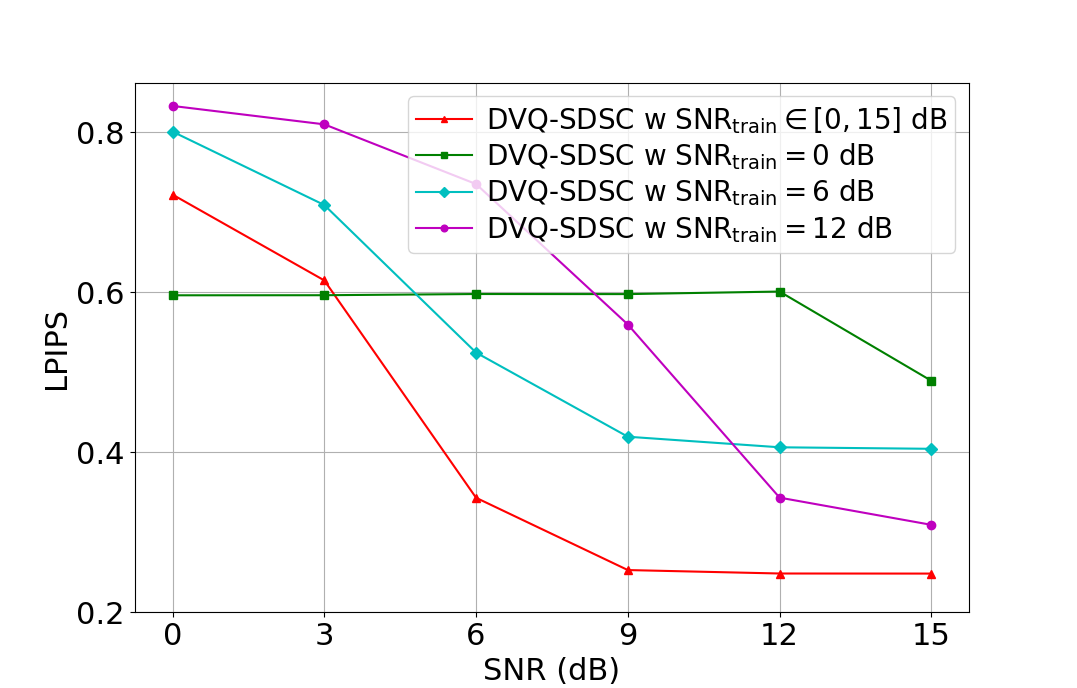}
\end{minipage}
}
\captionsetup{font={footnotesize}}
\caption{Objective performance comparison over AFDM with the 3GPP NTN-TDL-D channel. (a)-(c) Compare the proposed DVQ-SDSC against the JPEG+LDPC baseline in terms of PSNR, MS-SSIM, and LPIPS, respectively, where "Perfect CSI" and "Estimated CSI" denote the availability of ideal and estimated CSI at the receiver. (d)-(f) Ablation study on the training SNR regime, evaluating DVQ-SDSC trained over $[0,15]$ dB, fixed at $0$dB, $6$dB, and $12$dB, under the same three metrics. All DVQ-SDSC curves operate at $0.0625$ bpp, whereas the JPEG+LDPC baseline operates at $0.2095$ bpp.}
\label{figs}
\end{figure*}

\subsection{Experimental Setup} 

\textit{1) Training Environments and Evaluation Metrics} 

The proposed DVQ-SDSC framework is evaluated on FAIR1M [37], a large-scale fine-grained object recognition benchmark for high-resolution RSI. It contains over one million oriented bounding-box annotations across five categories and 37 sub-categories, with image sizes ranging from $1000\times1000$ to $10{,}000\times10{,}000$ pixels. Here, almost $5{,}000$ images are used for training and $500$ for testing.
During training, images are randomly cropped into $1024\times1024$ patches, whereas at test time they are divided into non-overlapping $4096\times4096$ images. The model is implemented in PyTorch and trained for 100 epochs using AdamW with an initial learning rate of $2\times10^{-4}$, weight decay $10^{-4}$, and a batch size of eight.  
All experiments are conducted on an NVIDIA RTX 4090 GPU.
Reconstruction quality is assessed via PSNR [38] for pixel-level accuracy, complemented by MS-SSIM [39] and LPIPS [40] for perceptual quality.

The bit compression ratio (BCR) $R_{\mathrm{BCR}}$ is defined as the ratio of the number of transmitted bits to that of the original image, serving as a measure of the required communication overhead. Since the proposed architecture is flexible enough to accommodate different BCRs, we configure the network parameters in this work to achieve $R_{\mathrm{BCR}}=\tfrac{1}{384}$.
Let $b_0$ denote the raw bits per channel per pixel ($b_0=8$), and let each quantization index be represented by $\log_2 K_T$ bits under fixed-length coding. Since SE-T and SE-B decimate the input by factors $\rho_T$ and $\rho_B$, respectively, the numbers of semantic feature vectors are $K_z=HW/\rho_T^{2}$ and $K_z'=HW/\rho_B^{2}$. The payload bits carried by the index sequence $\bm{I}^s$ are therefore
$L_{\rm bit} = (K_z + K_z')\log_2 K_T
= \left(\frac{1}{\rho_T^{2}} + \frac{1}{\rho_B^{2}}\right) HW \log_2 K_T $,
and $R_{\mathrm{BCR}}$ relative to raw image $\mathbf{S}$ is calculated by  
\begin{equation}
R_\mathrm{BCR} = \frac{L_{\rm bit}}{HWO\,b_0}
= \frac{\log_2 K_T}{O b_0}
\left(\frac{1}{\rho_T^{2}} + \frac{1}{\rho_B^{2}}\right).
\setcounter{equation}{26}
\label{eq26}
\end{equation}
Since \eqref{eq26} is independent of $H$ and $W$, the dual-branch codec is resolution-invariant.
For the configuration settings 
$\rho_T = \rho_B = 16$, $K_T = K_B = 256$, $O = 3$ and $b_0 = 8$,
we obtain $R_{\mathrm{BCR}} = \frac{1}{384} \approx 2.604 \times 10^{-3}$ and its corresponding bit per pixel (BPP) $R_{\rm bpp} = 0.0625$, which implies $131\,072$ indices, i.e., $1\,048\,576$~bit
($128$~KiB), for a $4096 \times 4096$ image against $48$~MiB of raw data. 
With a heterogeneous encoder-decoder architecture (0.52 M vs. 1.89 M parameters), the DVQ-SDSC offloads most computation to the ground station, minimizing satellite-side storage and computational burden while maintaining reconstruction fidelity. Coupled with low $R_{\mathrm{BCR}}$, this architecture realizes efficient end-to-end semantic transmission within stringent satellite resource constraints.

\textit{2) AFDM Symbol Structure and Channel Model} 

The AFDM system employs $N=1024$ DAFT-domain symbols with
$\Delta f=30$~kHz and a carrier frequency of $2$~GHz, yielding
$T=1/\Delta f=33.3$~$\mu$s and $\triangle t=T/N=32.55$~ns.
The chirp parameters are set to $c_1=[2(k_{\max}+N_v)+1]/(2N)=7/2048$
and $c_2=1/(20N)=1/20480$, with $k_{\max}=2$ and $N_v=1$.
Each frame comprises $N_p=32$ pilot symbols occupying $k\in\{0,\ldots,N_p-1\}$, a guard interval of $N_g=(l_{\max}+1)(2k_{\max}+2N_v)-1=11$ zero samples, $N_d=N-N_p-2N_g=970$
data symbols, and a trailing guard of $N_g=11$ zero samples, i.e., a frame efficiency of $N_d/N=94.73\%$. The observation window of length $N_T=N_p+N_g=43$ starts at
$\tilde{m}_s=k_{\max}+N_v-N_g=-8$ and is extracted via modulo-$N$
indexing. The fourth-order orthogonal amplitude modulation (QAM) is adopted. 
The satellite channel follows the 3GPP NTN-TDL-D profile [41], comprising line-of-sight (LOS) and non-LOS (NLOS) paths. The normalized maximum Doppler and delay spreads are $k_{\max}=2$ and $l_{\max}=1$, respectively, and the path delays are $\tau_i=\tau_{i,\mathrm{norm}}\cdot\tau_{\mathrm{spread}}$ for $i\in\{0,\ldots,L_p-1\}$, with $\tau_{\mathrm{spread}}\in[30,50]$ ns. Channel estimation is cast as sparse recovery with $\boldsymbol{\Phi}(\tilde{\mathbf{g}})\in\mathbb{C}^{43\times800}$ built on an $N_\tau=20$, $N_\nu=40$ delay-Doppler grid ($N_s=N_\tau N_\nu$), solved by OMP method.

\subsection{Performance of DVQ-SDSC Framework} 

We evaluate the end-to-end reconstruction performance of the proposed DVQ-SDSC framework over AFDM with standard 3GPP NTN-TDL-D channel. 
Unless otherwise specified, all DVQ-SDSC variants operate at $0.0625$ bpp.

\textit{1) Comparison with the Baseline Scheme} 

Figs. 5(a)-(c) compare DVQ-SDSC with the conventional JPEG+LDPC baseline under perfect and estimated CSI. Three conclusions follow. First, DVQ-SDSC consistently outperforms JPEG+LDPC over the entire 0-15dB SNR range in PSNR and MS-SSIM while attaining a substantially lower LPIPS score. This gain is achieved at only $0.0625$ bpp, which is much less than that of the $0.2095$ bpp consumed by the baseline, validating the effectiveness of the proposed semantic codec and VQ mechanism in preserving HF texture details under extreme bandwidth constraints. 
Second, the performance gap between perfect- and estimated-CSI cases remains marginal across all three metrics, confirming that the OMP-based sparse channel estimator reliably recovers CSI at the receiver and mitigates channel uncertainty. Consequently, DVQ-SDSC retains satisfactory reconstruction with estimated CSI, demonstrating the robustness of the joint AFDM waveform and OMP estimator under imperfect channel knowledge. Third, DVQ-SDSC's PSNR and MS-SSIM rise monotonically with SNR whereas LPIPS decreases steadily, confirming that the DJSCC-enabled semantic encoder-decoder successfully tracks varying channel conditions. In contrast, the JPEG+LDPC baseline
%which relies on separate source and channel coding, 
is fundamentally limited by its fixed compression-and-coding pipeline.

\begin{figure}
\centering
\subfigure[]{
  \label{fig:subfig:a}
   \includegraphics[width = 0.39\textwidth]{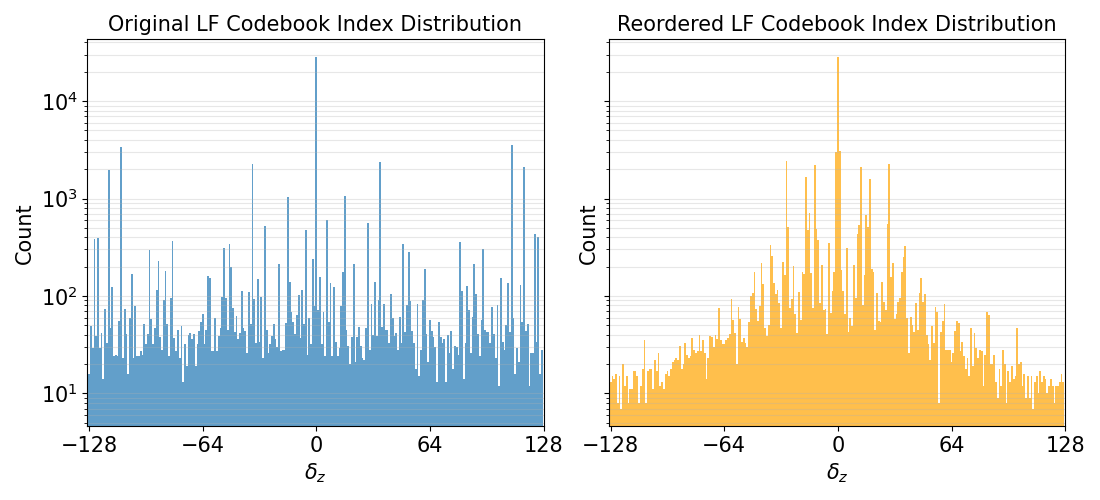}}
\hspace{1in}
\subfigure []{
 \label{fig:subfig:b}
 \includegraphics[width = 0.39\textwidth]{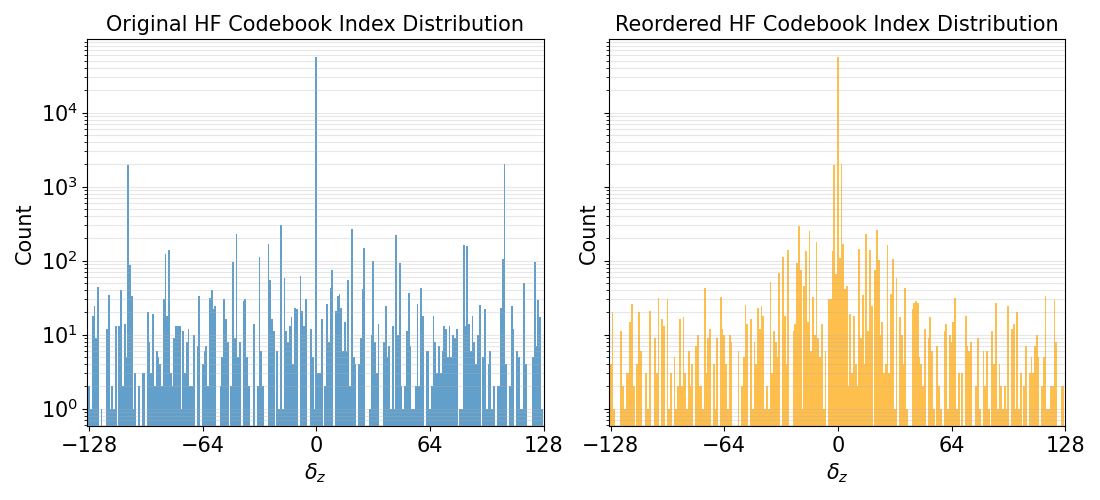}}
 \hspace{1in}
\subfigure []{
 \label{fig:subfig:b}
 \includegraphics[width = 0.39\textwidth, height =0.15\textheight]{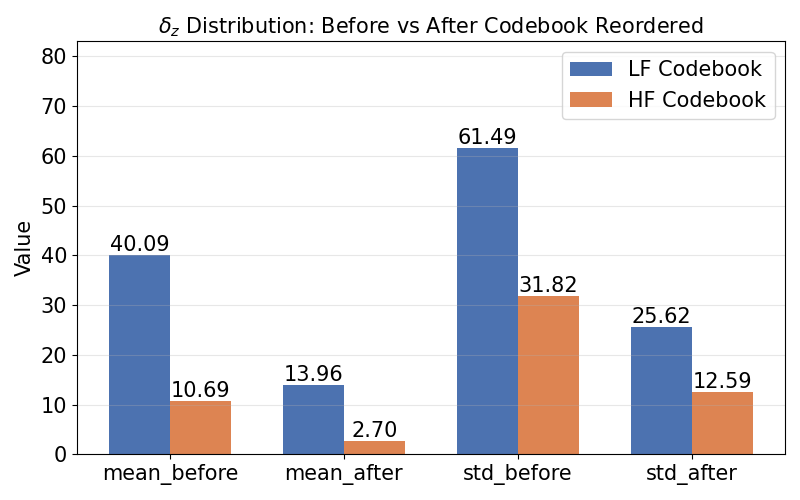}} 
\captionsetup{font={footnotesize}}
\caption{Empirical evaluation of the DPCM-aware PCA codebook reordering.
(a) Distribution of the folded residual $\delta_z$ for the LF index sequence, obtained without (blue) and with (orange) the
proposed reordering $\bm{\pi}$.
(b) Corresponding distribution of $\delta_z$ for the HF index sequence.
(c) Mean and standard deviation of $\delta_z$ for LF and HF branches
before and after reordering.
} 
\label{fig:subfig}
\end{figure}

\textit{2) Impact of the SNR Adaptive Module} 

To assess the proposed SNR-adaptive mechanism, Figs. 5(d)-(f) compare models with and without it. Four configurations are evaluated: the full SNR-adaptive model trained over the entire $[0,15]$dB interval, and three non-adaptive baselines trained and tested at fixed $0$, $6$, and $12$dB. The latter are optimized for a single predetermined channel condition and thus lack adaptivity. As evidenced by the curves, the full model achieves the most balanced performance across the entire test range, remaining competitive in PSNR, MS-SSIM, and LPIPS at both low- and high-SNR endpoints. In contrast, the fixed-SNR models exhibit a pronounced trade-off: (i) the $12$dB model performs well at high SNR but degrades noticeably at $0$-$3$dB, while the $0$dB model shows the opposite behavior; (ii) the $6$dB model compromises between the two yet still falls short, particularly in perceptual quality. This gap directly demonstrates that the model remains tied to its training SNR and generalizes poorly to unseen channel states without the SNR-adaptive module.

%These results substantiate the effectiveness of the proposed SNR-adaptive module, which dynamically adjusts the semantic representation and channel adaptation according to the instantaneous SNR. By exposing the network to the full $[0,15]$-dB interval during training, the module learns a continuous and channel-aware mapping that bridges diverse SNR conditions, thereby endowing the model with robust generalization across the entire range. The superiority of the full-adaptive configuration over the fixed-SNR baselines thus confirms that the gain originates from the adaptive mechanism itself, rather than merely from the training data coverage.

%In summary, the simulation results in Fig. 5 demonstrate that: (i) DVQ-SDSC significantly outperforms the conventional JPEG+LDPC baseline at a much lower bpp; (ii) the impact of CSI estimation error is effectively suppressed by the OMP-based sparse channel estimator; and (iii) the proposed SNR-adaptive module achieves consistently superior robustness and generalization over non-adaptive fixed-SNR models. These findings confirm the validity of the proposed dual-branch architecture and the associated joint optimization framework for high-resolution RSI transmission over LEO satellite channels.

\begin{figure*}[t] %

\subfigure[]
{
\begin{minipage}[c]{5.6cm}
\centering
\includegraphics[width=1.1\linewidth]{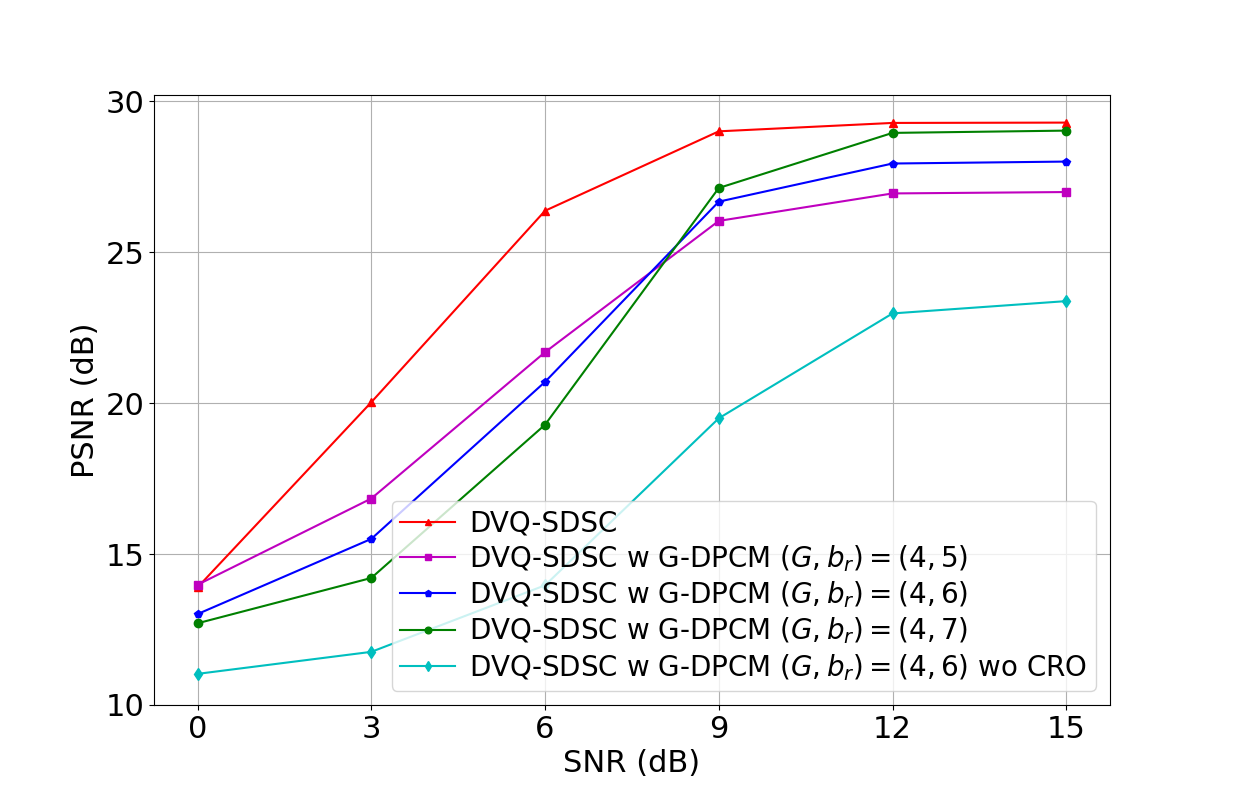}
\end{minipage}
}
\subfigure[]
{
\begin{minipage}[c]{5.6cm}
\centering
\includegraphics[width=1.1\linewidth]{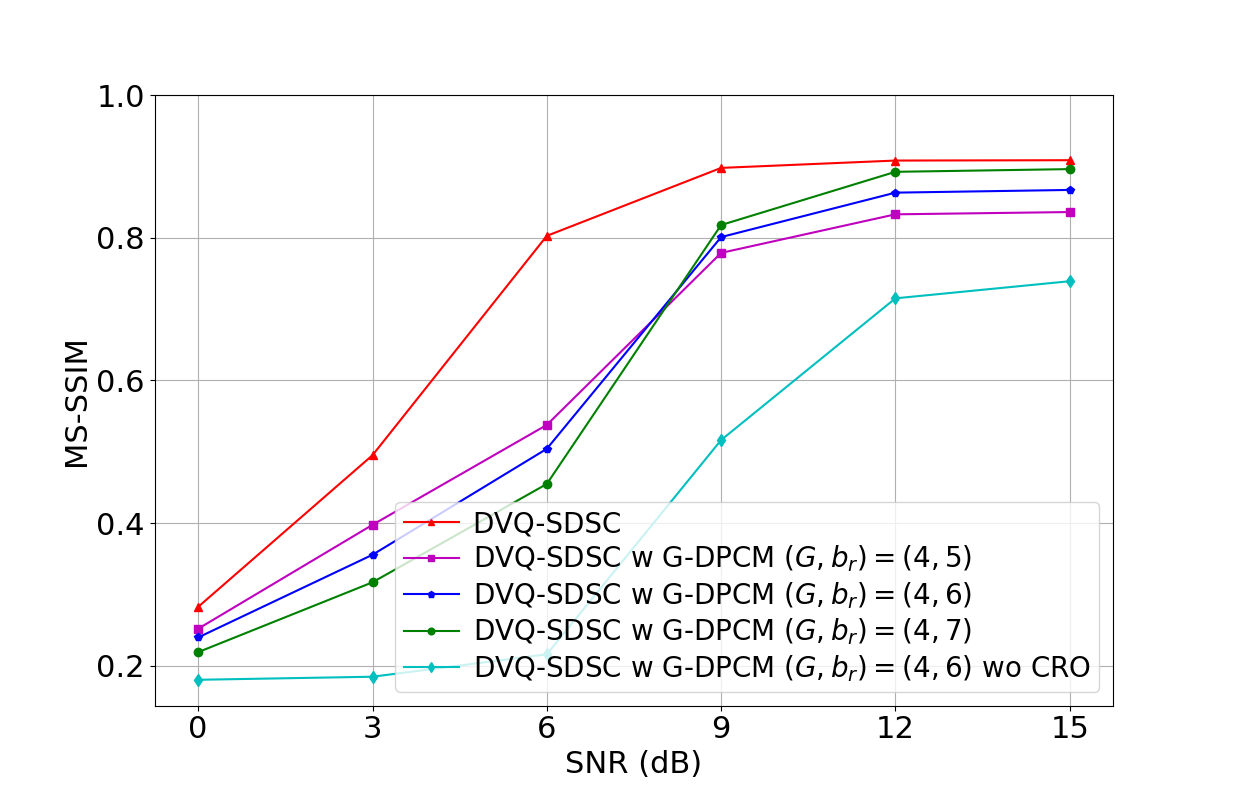}
\end{minipage}
}
\subfigure[]
{
\begin{minipage}[c]{5.6cm}
\centering
\includegraphics[width=1.1\linewidth]{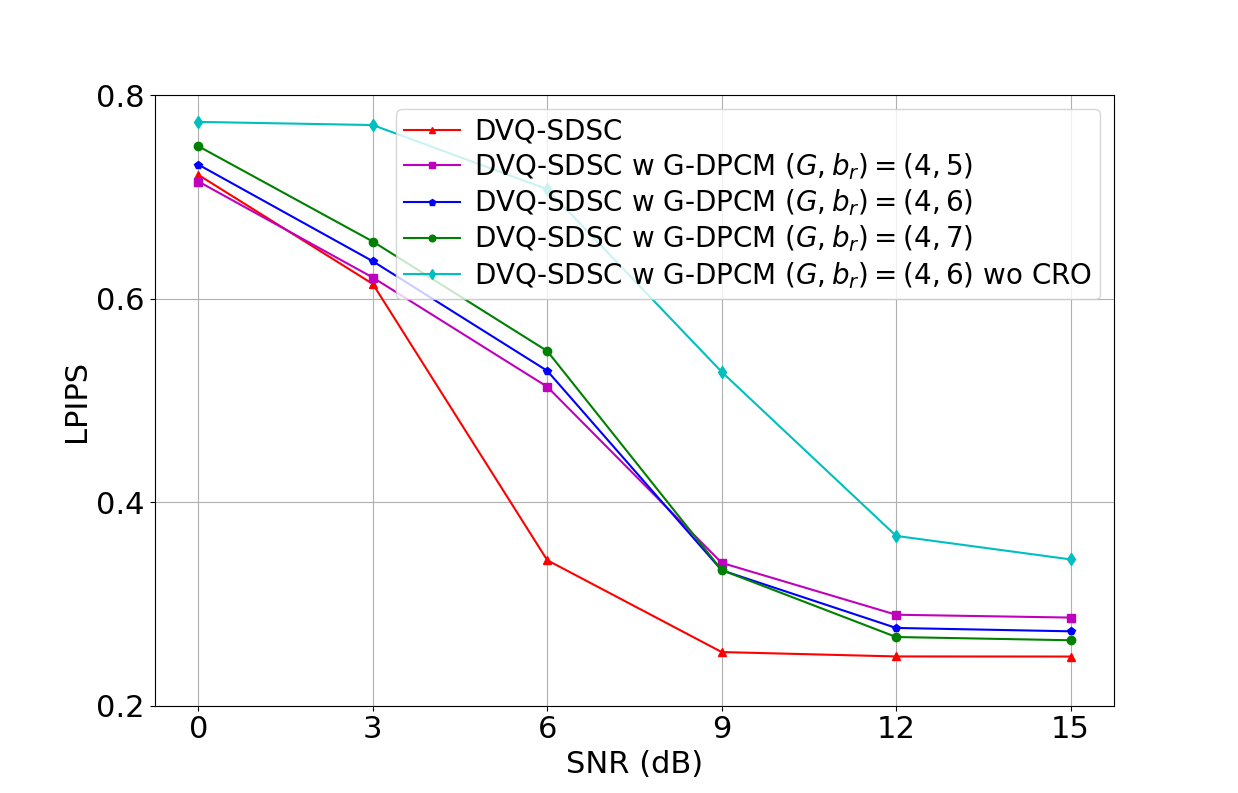}
\end{minipage}
}
\captionsetup{font={footnotesize}}
\caption{Performance comparison between the proposed DVQ-SDSC and its variant with the G-DPCM codec-based index compression and transmission, evaluated over AFDM with 3GPP NTN-TDL-D channel. %at a fixed compression rate of $0.0625$ bpp. 
(a) PSNR, (b) MS-SSIM, and (c) LPIPS as functions of the wireless SNR in the range $0$-$15$ dB. 
%``DVQ-SDSC wo G-DPCM'' (red) transmits every quantized VQ index as a fixed-length $b_a=\lceil\log_2 K\rceil=8$ bits. The remaining three curves employ the proposed G-DPCM codec with group size $G=4$ and residual bit widths $b_r\in\{5,6,7\}$, colored purple, blue, and green, respectively; their index-overhead savings relative to fixed-length coding are predicted by $\rho_G=Gb_a/[b_a+(G-1)b_r]$ in \eqref{eq:rhoG}, i.e., $28.12\%$, $18.75\%$, and $9.38\%$. The results jointly demonstrate the rate--distortion--perception trade-off enabled by the DPCM-aware PCA reordering analyzed in Fig.~\ref{fig:6}.
}
\label{figs}
\end{figure*}

\subsection{Performance of Codebook Reordering and G-DPCM}

%To assess the residual concentration asserted in \textit{Lemma 1}, we conduct a
%comprehensive simulation analysis on the empirical distribution of the folded
%residual $\delta_z$ defined in (29), together with its associated first- and second-order statistics. For each $4096\times4096$ test image, both SE-T and SE-B produce a $256\times128$ index map,
%i.e., $65{,}536$ indices per branch. Restricting attention to horizontally
%adjacent pairs yields $65{,}280$ residual samples per branch per image, and all
%statistics below are aggregated over the complete test set. The blue bars/curves
%correspond to the trained codebook in its original ordering,
%whereas the orange bars/curves employ the permutation $\bm{\pi}$ returned by
%\textbf{Algorithm 2}. Since $K_T=K_B=256$, the residual takes values in
%$[-\lfloor K/2\rfloor,\lfloor K/2\rfloor-1]=[-128,127]$.

To assess the residual concentration asserted in \textit{Lemma 1}, we analyze the empirical distribution of the folded residual $\delta_z$ together with its first- and second-order statistics. For each $4096\times4096$ test image, SE-T and SE-B each produce a $256\times128$ index map. 
Since $K_T=K_B=256$, the residual lies in $[-128,127]$.
The blue bars/curves correspond to the trained codebook in its original ordering, whereas the orange ones employ the permutation $\bm{\pi}$ returned by \textbf{Algorithm 2}.
Figs. 6(a) and 6(b) show that codebook reordering markedly contracts the residual mass toward the origin in both LF and HF branches. Under the original ordering, non-zero residuals disperse over essentially the entire ring as expected. Since the trained codewords carry no topological ordering, two spatially adjacent latents may map to codewords with arbitrarily distant indices. After reordering, the distribution becomes sharply peaked and unimodal with most non-zero mass confined to a narrow interval around zero, which is precisely predicted by \textit{Lemma 1(i)}. 
Because $\bm{\pi}$ is a
bijection, it preserves the event that two spatially adjacent latents are
quantised to the same codeword, and hence the number of pairs with $\delta_z=0$ is identical for both orderings. 
The gain from \textbf{Algorithm 2} is therefore the contraction of the non-zero residual support, as evidenced by Fig. 6(c), which shows a pronounced reduction in both mean absolute value and standard deviation.

To isolate the end-to-end impact of the proposed G-DPCM codec (shown as \textbf{Algorithm 3}), we assess reconstruction quality as a function of both the channel conditions and the residual bit width $b_r$. For a fair comparison, all curves in Fig. 7 share the same base compression rate $R_{\mathrm{bpp}}=0.0625$ bpp, and only the overhead introduced by index transmission is varied. Setting $G=4$ and $b_a=8$, the group packing factor becomes $\rho_G=Gb_a/[b_a+(G-1)b_r]$, so a smaller $b_r$ produces a larger $\rho_G$ and hence fewer transmitted bits. The three G-DPCM configurations correspond to $b_r\in\{5,6,7\}$, whose theoretical rate savings are $28.125\%$, $18.75\%$, and $9.375\%$, respectively, relative to fixed-length index coding.

The results of Fig. 7 yield four conclusions. (i) Among the evaluated schemes, the fixed-length reference (red) achieves the highest fidelity, while the three G-DPCM curves degrade monotonically with decreasing $b_r$. (ii) This degradation is gradual rather than abrupt, and the default $b_r=6$ (blue) trails the red curve only slightly despite transmitting $18.75\%$ fewer index bits. Since the DPCM-aware PCA reordering $\boldsymbol{\pi}$ concentrates the folded residual $\delta_z$ about the origin, the majority of samples stay within the clipping interval, so reducing $b_r$ sacrifices little reconstruction quality. (iii) The $(G,b_r)=(4,5)$ configuration (green) is particularly revealing: at low SNR, its small group size $G=4$ increases anchor-bit exposure to channel errors, while its coarse $b_r=5$ quantization halves the clipping interval relative to $b_r=6$, jointly degrading residual-decoding robustness. (iv) The ablated $(G,b_r)=(4,6)$ variant without codebook reordering (wo CRO, cyan) consistently lies below the G-DPCM curves, demonstrating that PCA-based codebook reordering is essential for residual concentration.

Moreover, we relate the residual contraction in Fig. 6 to the coding gain of
\textit{Lemma~1(ii)}.
Because G-DPCM assigns a fixed $b_r$-bit field to every residual, the reordering
$\bm{\pi}$ itself does not change the bit rate, and it only determines the residual
dynamic range $B_{\max}$ that this field must cover. With $b_a=8$, $b_r=6$, and $G=4$, it gives
$18.75\%$ bit saving over fixed-length coding, independent of the residual
statistics. Equivalently, both index maps of a $4096\times4096$ image require
$2L_s b_a=1.049$~Mbit without G-DPCM versus
$2\lceil L_s/G\rceil\,[b_a+(G-1)b_r]=0.852$~Mbit with it, corresponding to
$384\times$ and $473\times$ compression relative to 8-bit RGB. \textbf{Algorithm 3} is clipping-free only if $2^{b_r}-1\geq B_{\max}$, i.e., the
offset-clip in \eqref{eq24} spans $[-32,31]$ for $b_r=6$. 
The residual concentration achieved by PCA reordering eliminates the need for a wider clipping field and preserves a larger $\rho_G$, thereby enabling the gradual rate-distortion-perception trade-off in Fig. 7.

\begin{figure*}[htbp]
\centering
\resizebox{0.95\textwidth}{0.28\textheight}{\includegraphics{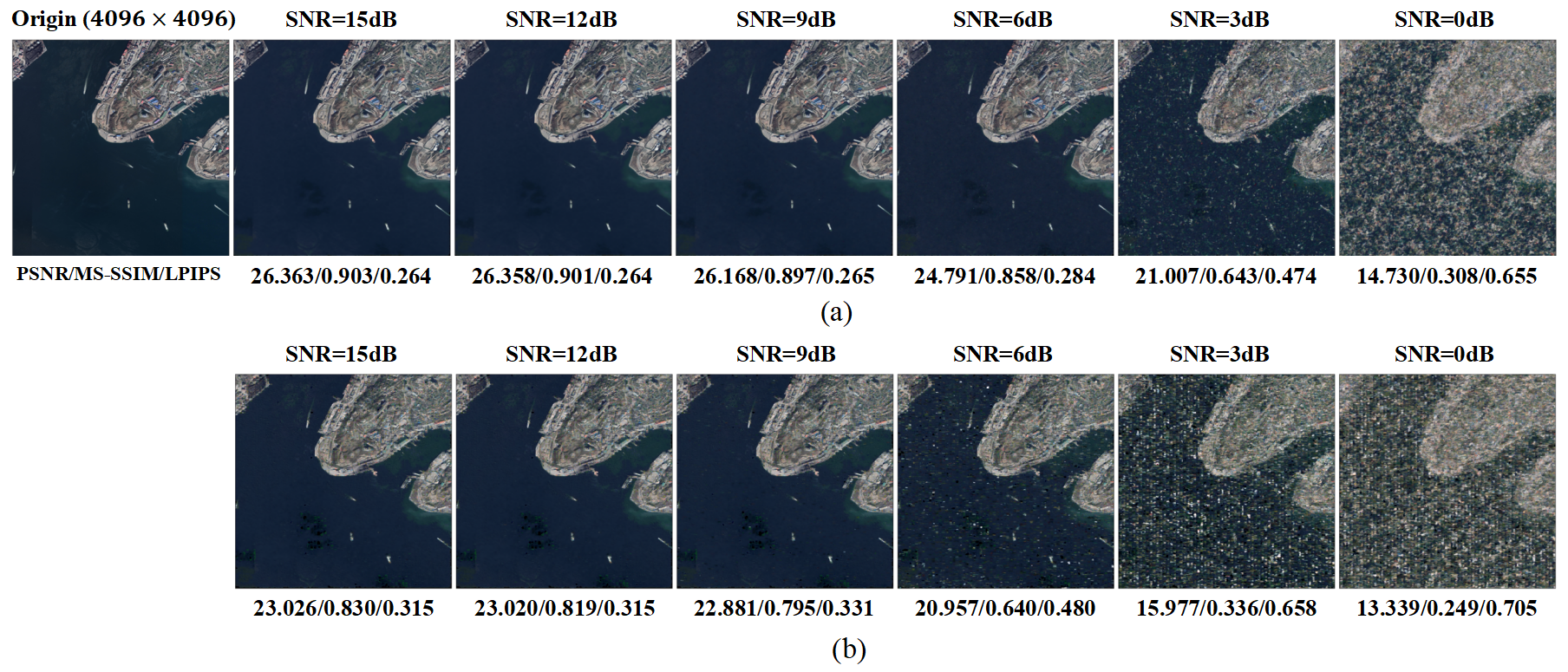}}
\captionsetup{font={footnotesize}}
\caption{
Two comparative visual examples between the proposed DVQ-SDSC and its variant incorporating the G-DPCM codec, both operating at $(G, b_r)=(4,6)$ and $R_{\mathrm{bpp}}=0.0625$. The reconstructed images are displayed from left to right in descending order of SNR, ranging from 15 dB to 0 dB. (a) DVQ-SDSC, and (b) DVQ-SDSC with the G-DPCM codec; the PSNR, MS-SSIM, and LPIPS scores are annotated below each image.
}
\label{fig8}
\end{figure*}

The visual comparisons in Fig. 8 corroborate the quantitative findings in Fig. 7. At high SNR (15-9 dB), both the fixed-length DVQ-SDSC (Fig. 8(a)) and its G-DPCM variant (Fig. 8(b)) reconstruct visually faithful RSIs, preserving coastline geometry and land-sea boundaries with only subtle differences. However, both schemes accumulate noise and granular artifacts with fine textures and edges progressively blurring below 6 dB.
%with fine textures and edges progressively blurring.
%at 0 dB, the images are nearly submerged in speckle. 
A closer comparison at identical SNR and bit allocation ($R_{\mathrm{bpp}}=0.0625$, $(G,b_r)=(4,6)$) shows that the G-DPCM variant retains quality comparable to the fixed-length codec at 15-9 dB, while degrading markedly more gracefully at 6-0 dB without abrupt structural collapse. 
This directly reflects the residual contraction in \textit{Lemma 1}: the DPCM-aware PCA reordering $\boldsymbol{\pi}$ concentrates the folded residual $\delta_z$ about the origin, confining most samples within the clipping interval; consequently, reducing $b_r$ sacrifices little reconstruction quality while isolating clipping errors locally within each group. In summary, G-DPCM reduces index-transmission overhead by $18.75\%$ with no latent perceptual penalty, particularly at high SNR.

\section{Conclusion}

In this paper, we proposed DVQ-SDSC, a VQ-aided dual-branch digital semantic communication framework for high-resolution RSI transmission over AFDM-based satellite channels. First, we introduced an asymmetric codec that decouples HF texture residuals from LF structural semantics via bicubic extraction, multi-scale fusion, channel attention, and VQ quantization, followed by gated fusion and SNR-adaptive reconstruction. This hierarchical design overcomes the severe HF loss of conventional semantic codecs under extreme bandwidth constraints.
Second, we developed a index compression scheme that combines PCA-aided codebook reordering with G-DPCM. By contracting the folded residual $\delta_z$ toward the origin, it enables a graceful rate-distortion-perception trade-off and localizes clipping and channel errors within groups.
Third, extensive FAIR1M experiments over 3GPP NTN-TDL-D channels show that DVQ-SDSC surpasses JPEG+LDPC across PSNR, MS-SSIM, and LPIPS at 0.0625 bpp, which is less than one third of the 0.2095 bpp baseline while remaining robust under estimated CSI.
Finally, the heterogeneous architecture (0.52M/1.89M parameters) offloads computation to the ground station, while G-DPCM reduces index overhead with negligible perceptual penalty and no retraining, making it well suited to satellite-to-ground resource constraints.

\end{document}